\documentclass[aip,reprint]{revtex4-1}

\usepackage{graphicx}%
\usepackage{multirow}%
\usepackage{amsmath,amssymb,amsfonts}%
\usepackage{siunitx}
\usepackage{hyperref}

\draft 

\makeatletter
\def\@email#1#2{%
 \endgroup
 \patchcmd{\titleblock@produce}
  {\frontmatter@RRAPformat}
  {\frontmatter@RRAPformat{\produce@RRAP{*#1\href{mailto:#2}{#2}}}\frontmatter@RRAPformat}
  {}{}
}%
\makeatother
\begin{document}


\title{Advanced fabrication process for particle absorbers of highly pure electroplated gold for microcalorimeter applications} 


\author{Michael Müller}
\email[]{michael.mueller2@kit.edu}
\affiliation{Institute of Micro- and Nanoelectronic Systems (IMS), Karlsruhe Institute of Technology (KIT), Hertzstraße 16, 76187 Karlsruhe, Germany}

\author{Ria-Helen Zühlke}
\affiliation{Institute of Micro- and Nanoelectronic Systems (IMS), Karlsruhe Institute of Technology (KIT), Hertzstraße 16, 76187 Karlsruhe, Germany}

\author{Sebastian Kempf}
\affiliation{Institute of Micro- and Nanoelectronic Systems (IMS), Karlsruhe Institute of Technology (KIT), Hertzstraße 16, 76187 Karlsruhe, Germany}
\affiliation{Institute for Data Processing and Electronics (IPE), Karlsruhe Institute of Technology (KIT), Hermann-von-Helmholtz-Platz 1, 76344 Karlsruhe, Germany}


\date{\today}

\begin{abstract}
Magnetic microcalorimeters (MMCs) have become a key technology for applications requiring outstanding energy resolution, fast signal rise time and excellent linearity. MMCs measure the temperature rise upon absorption of a single particle within a particle absorber by using a paramagnetic temperature sensor that is thermally coupled to the absorber. The design and fabrication of the particle absorber is key for excellent detector performance. Here, we present a microfabrication process for free-standing particle absorbers made of two stacked and independently electroplated high-purity Au layers. This enables, for example, embedding of radioactive sources within the absorber for realizing a $4\pi$ detection geometry in radionuclide metrology or preparing detector arrays with variable quantum efficiency and energy resolution as requested for future applications in high energy physics. Due to careful optimization of photoresist processing and electroplating parameters, the Au films are of very high purity and very high residual resistivity ratio values above 40, allowing for fast internal absorber thermalization.
\end{abstract}


\maketitle 

\section{Introduction}
\begin{figure*}
    \centering
    \includegraphics[width=0.65\textwidth]{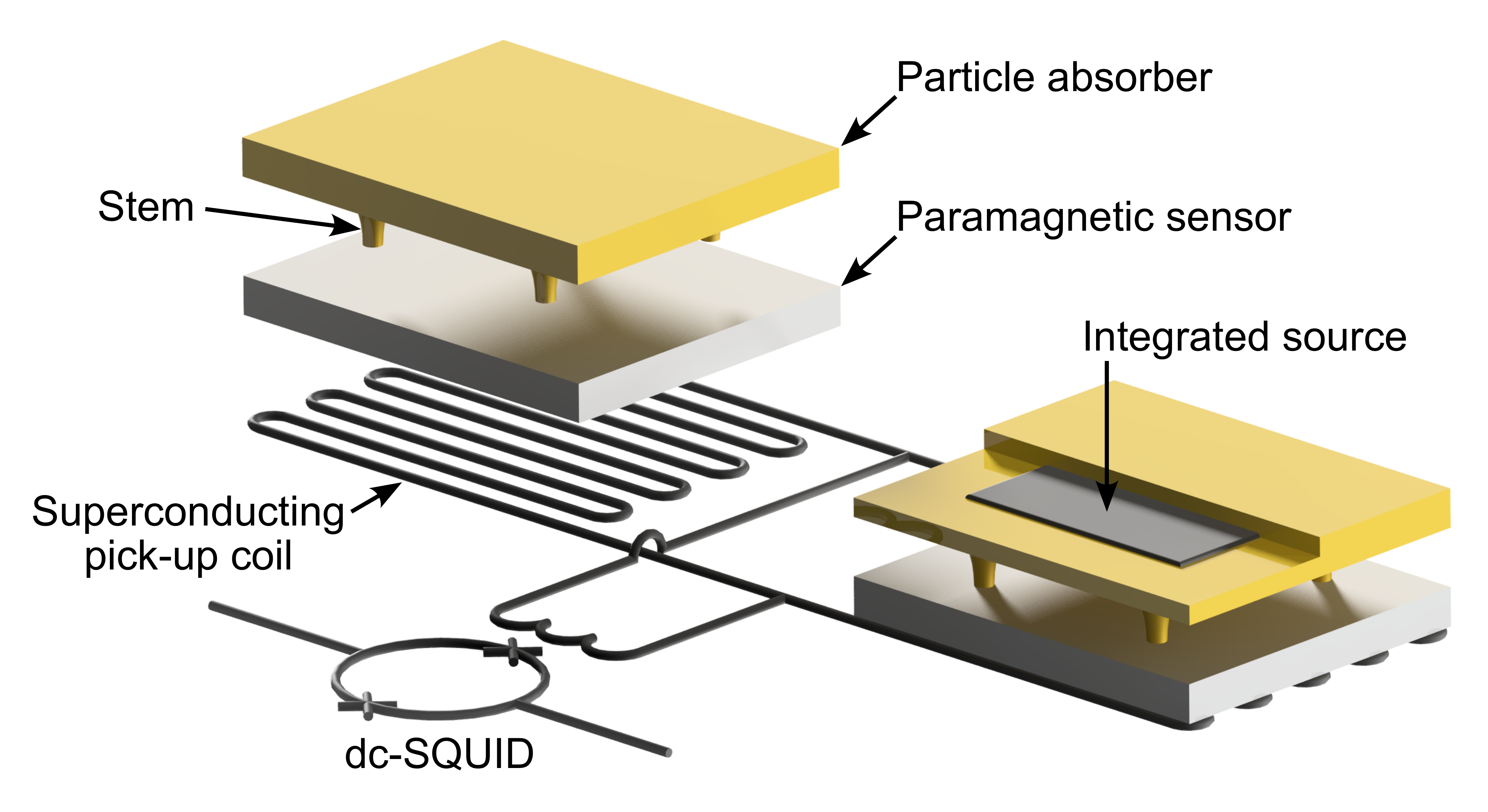}
    \caption{\label{fig:MMC_Render}Three-dimensional illustration  of a microfabricated MMC comprising two superconducting meander-shaped pickup coils that are connected in parallel with the input coil of the readout dc-SQUID. As indicated at the right detector pixel, the particle absorbers allow integrating a radioactive source into the absorber to yield a $4\pi$ detection geometry.}
\end{figure*}

Magnetic microcalorimeters (MMCs) are state-of-the-art cryogenic detectors for energy-dispersive single particle detection and have, in the recent years, become an extraordinary powerful tool for a variety of applications \cite{Fleischmann2005, Kempf2018}. They make use of the properties of a paramagnetic temperature sensor placed in a weak magnetic field to yield a very strong dependence of the sensor magnetization on temperature. Typical sensor materials are Au:Er or Ag:Er operated at temperatures below \SI{100}{mK}. The sensor is in tight thermal contact to a particle absorber and weakly coupled to a heat bath. Upon absorption of a particle with energy $E$, the detector temperature rises by $\Delta T = E/C_{\mathrm{tot}}$, with $C_{\mathrm{tot}}$ denoting the total heat capacity of the detector. The magnetic flux change resulting from the change in magnetization is read out with high precision using a direct current superconducting quantum interference device (dc-SQUID). Fig.~\ref{fig:MMC_Render} depicts a three-dimensional illustration of state-of-the-art microfabricated MMC comprising two superconducting meander-shaped pickup coils that are connected in a gradiometric configuration to reduce the influence of temperature fluctuations of the heat bath as well as fluctuations of external background fields. The pick-up coils provide a magnetic field bias to the overlying sensor layers by storing a persistent current and pick up the temperature-induced magnetization changes. On top of the sensors, the particle absorbers are placed, typically free-standing on a few stems, to avoid athermal phonon loss during internal absorber thermalization \cite{Kempf2018}.

Recent experiments demonstrate that MMCs achieve outstanding energy resolution as good as \SI{1.25}{eV} for photon energies of $5.9\,\mathrm{keV}$ \cite{Krantz2024}. Furthermore, MMCs provide a very fast signal rise time in the order of \SI{100}{ns} as well as very good linearity of the detector response \cite{Fleischmann2005, Rotzinger2008}. Current MMC applications include the search for dark matter \cite{Krosigk2023}, direct neutrino mass determination \cite{Gastaldo2017}, X-ray spectroscopy \cite{Sikorsky2020, Hengstler2015, Pfaefflein2022, Herdrich2024}, search for neutrinoless double beta decay \cite{Alenkov2019DoubleBetaDecay, Kim2023DoubleBetaDecay}, X-ray astronomy \cite{Stevenson2019HydraMMC, Devasia2022LynxMMC} and many more. A specific example in radionuclide metrology is the EMPIR project ``PrimA-LTD - Towards new primary activity standardisation methods based on low-temperature detectors'' that used MMC technology to demonstrate activity standardization with unprecedentedly low uncertainties as well as high resolution decay energy spectrometry on low-energy decaying radionuclides \cite{Mueller2024PrimALTD, Kossert2022} More information about the project can be found on the official PrimA-LTD website: https://prima-ltd.net/.

The design and fabrication process of the particle absorbers are very important for any of these applications and play key roles for well-performing MMCs. There are three parameters to be optimized: 1) a large quantum efficiency for high particle absorption probability, 2) a low heat capacity for a large detector signal and 3) a high thermal conductivity for fast internal thermalization of the absorber. To achieve a quantum efficiency close to \SI{100}{\percent}, the thickness of the particle absorber is determined taking into account the energy of the particles to be measured as well as the absorber material, which sets a lower limit to the detector's heat capacity $C_{\mathrm{tot}}$. Gold is an established absorber material as its high atomic number reduces the required absorber thickness and therefore heat capacity. Furthermore, its high thermal conductivity provides fast internal absorber thermalization to prevent position-dependencies of the signal shape \cite{FigueroaFeliciano2008}. To prevent energy loss in the early stage of thermalization due to athermal phonons escaping the detector through the substrate \cite{Kozorezov2013}, stems with height and diameter of several \SI{}{\micro\m} are typically introduced, on which the absorbers are free-standing (see Fig.~\ref{fig:MMC_Render} and \cite{Kempf2018, Fleischmann2009}). This increases the phonon interaction length in the absorber by drastically reducing the contact area to the sensor, enhancing the achievable energy resolution but making the fabrication process more challenging.

In this work, we present a microfabrication process for free-standing absorbers that are composed of two individually electroplated Au layers. This allows on the one hand the fabrication of particle absorbers with different thicknesses on the same wafer, e.g. by covering only a fraction of the absorbers with a second absorber half. The usage of such devices is strongly considered in the context of the microcalorimeter detectors for high-energy physics (HEP) experiments \cite{Bass2024}, for example in the search for axions \cite{IAXO2013}. On the other hand, the presented process enables the integration of a radioactive source into the absorber to yield a $4\pi$ detection geometry as schematically depicted in Fig.~\ref{fig:MMC_Render}. Experiments that use integrated sources in low-temperature calorimeters are, for example, ECHo \cite{Gastaldo2017} and HOLMES \cite{Alpert2015HOLMES}, that aim for direct neutrino mass determination by precise measurement of the $^{163}$Ho electron capture spectrum. In these experiments, the second absorbers halves of a few \SI{}{\micro\m} thickness are presently deposited on top of the source by sputtering or evaporation. While such a procedure is suitable for comparably thin absorber layers, the deposition of thicker layers leads to an expensive material usage and introduces further challenges like photoresist stability. In the framework of the EMPIR project PrimA-LTD, for instance, one goal was the high-resolution, high statistics decay energy spectrometry on the electron capture decaying nuclide $^{55}$Fe. This required a Au thickness of \SI{12}{\micro\m} all around the source to achieve the a detection efficiency of \SI{99.99}{\percent}, making electroplating the much preferred deposition technique.

Electroplating of all absorber layers not only minimizes the material costs but further provides Au with high thermal conductivity in the entire absorber volume. This is because electroplating of Au is known to produce films with greatly improved purity as compared to other deposition techniques such as evaporation or sputtering. A typically used qualitative measure for the sample purity is the residual resistivity ratio (RRR), that is defined as the ratio of room temperature resistivity and resistivity at \SI{4.2}{K}: $\mathrm{RRR}=\rho_{\SI{300}{K}}/\rho_{\SI{4.2}{K}}$. This is because the mean free path of the electrons at room temperature is dominated by electron-phonon scattering, whereas at low temperatures, it is set by scattering on lattice defects like grain boundaries and on any impurities. In very pure films, the geometric limitation of the sample sets an upper limit to the electron mean free path and therefore to the RRR that can be achieved \cite{Brown2008}. According to the Wiedemann-Franz law, the thermal conductivity of a metal is linked to its electrical conductivity, therefore, a high RRR value is an indicator for high thermal conductivity at low temperatures \cite{Irwin2005} and it should to be maximized to suppress any position dependence of the detector signal.


\begin{figure*}
    \includegraphics[width=.8\textwidth]{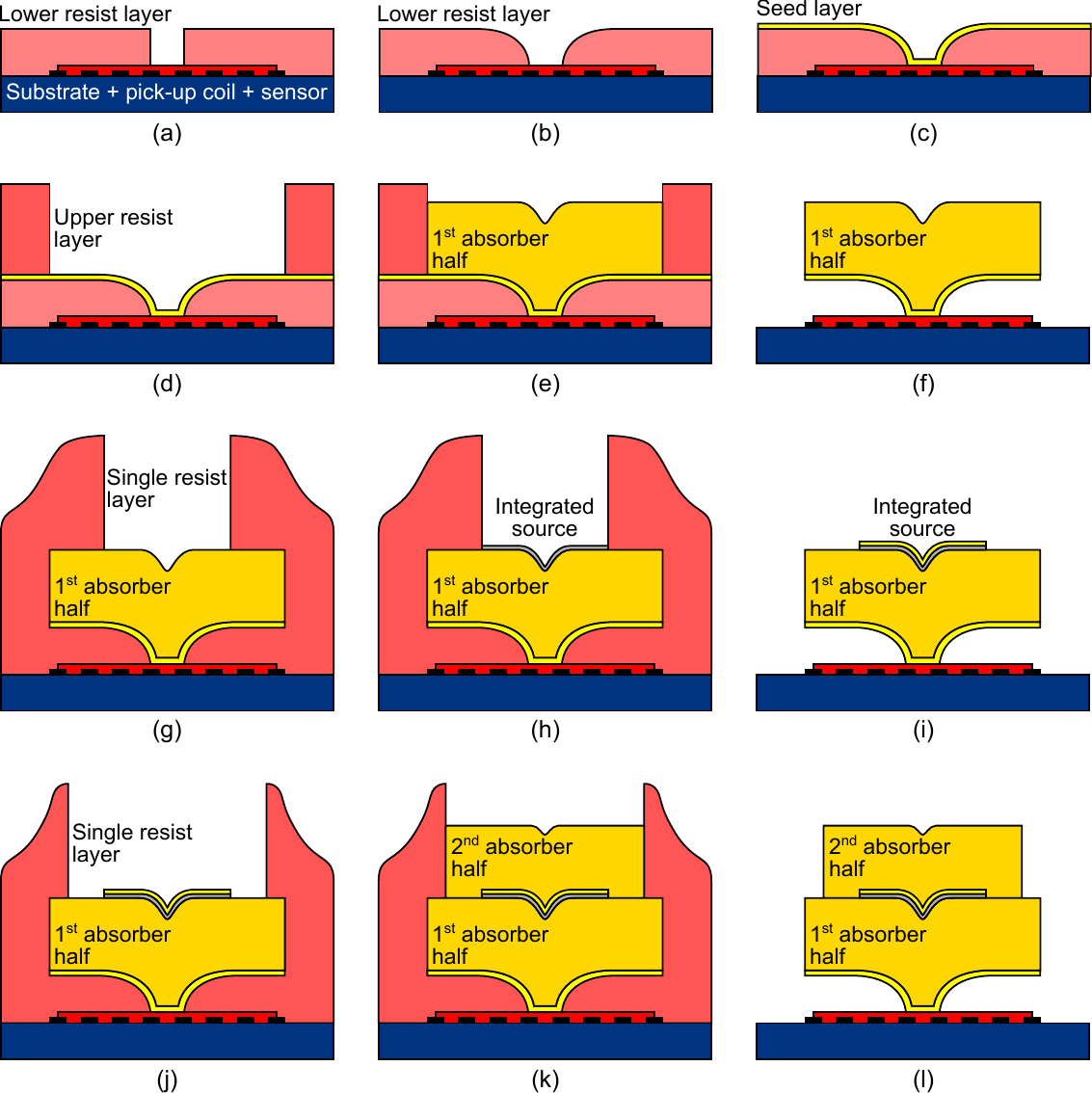}
    \caption{\label{fig:FabProcess}Schematic illustration of the individual fabrication steps for particle absorbers consisting of two independently electroplated Au layers with an integrated radioactive source in between. (a) Structuring a photoresist layer with holes for the absorbers stems onto the detector wafer. (b) Thermal reflow of the photoresist for rounded edges. (c) Deposition of a thin Au seed layer onto the entire wafer. (d) Structuring of the upper photoresist layer that defines the absorber size. (e) Electroplating of the first absorber half. (f) Releasing the absorber by removing supporting photoresists and seed layer. (g) Structuring of a photoresist that defines the size of the integrated source. (h) Integration of the source, e.g. by drop deposition, ion-implantation, etc. (i) Capping the integrated source with thin Au layer and removal of the photoresist. (j) Structuring of a photoresist layer that defines the size of the second absorber half. (k) Electroplating of the second absorber half. (l) Removal of the photoresist.}
\end{figure*}

\section{Overview of the absorber fabrication and source embedding process}
The fabrication of fully microfabricated particle absorbers with $4\pi$ detection geometry is split into three main steps: 1) Electroplating of the first absorber half that is virtually free-standing and mechanically supported by only a few small-diameter stems, 2) source deposition on top or source integration into the first absorber half and 3) electroplating of the second absorber half on top. In the following, we give an overview on the fabrication procedure and outline challenges we have faced during process development. Various details on photoresist processing, Au electroplating, substrate treatment etc. are then given in the sections below.

Fig.~\ref{fig:FabProcess} depicts the sequence of fabrication steps that we have developed to realize fully microfabricated particle absorbers with $4\pi$ detection geometry. The fabrication of the first absorber half follows essentially the fabrication process that is used for fabricating MMC-based detectors with free-standing X-ray absorbers \cite{Fleischmann2009}. It starts with depositing a positive-tone photoresist layer on top of the detector wafer which is finalized except for the particle absorbers. This resist layer is patterned such that only the parts of the resist layer where the absorber is supposed to contact the temperature sensor, i.e. where the absorber stems are located, are removed during resist developing (see Fig.~\ref{fig:FabProcess}a). The photoresist is then thermally reflowed to achieve a smoothened / rounded resist profile (see Fig.~\ref{fig:FabProcess}b). This enhances the stability of the absorbers and ensures that the subsequently sputter-deposited thin Au layer is continuous and not intersected by sharp resist edges (see Fig.~\ref{fig:FabProcess}c). This layer acts as a seed layer to which the cathode of the electroplating setup is later electrically connected. Next, a second positive-tone photoresist layer is spun onto the wafer. This layer is patterned such that molds defining the absorber area are formed during the resist development process (see Fig.~\ref{fig:FabProcess}d). In the next step, the molds are filled with Au by electroplating up to the target absorber thickness (see Fig.~\ref{fig:FabProcess}e). The latter is adjusted by varying the time of electrodeposition. Finally, the absorber is released by removing both resist layers as well as the parts of the seed layer that are not underneath the particle absorbers (see Fig.~\ref{fig:FabProcess}f). To achieve straight edges of the electroplated particle absorbers, the second photoresist layer must be significantly thicker than the target absorber thickness. Such photoresists typically require specific handling and processing parameters. The inherent interplay with the lower first resist layer, however, significantly constrains the usable process parameter space and makes a careful process optimization crucial.

After fabrication of the first absorber half, the radioactive source to be embedded into the absorber is deposited. This is done by any available radioactive source deposition technique, e.g. drop deposition and subsequent drying the deposited droplets \cite{Bockhorn2020}, electroplating \cite{Loidl2018}, or much more advanced, ion implantation \cite{Schneider2016, Niemeyer2024}. In any case, the source is deposited by spin-coating a photoresist layer and patterning this layer such that the part of the first absorber half where the source should be deposited is removed during photoresist development (see Fig.~\ref{fig:FabProcess}g). After source deposition (see Fig.~\ref{fig:FabProcess}h), the source is encapsulated by a thin, sputter-deposited or evaporated Au layer to protect the deposited source against subsequent steps of absorber fabrication and the photoresist is removed (see Fig.~\ref{fig:FabProcess}i). When ion-implantation is the method of choice, the additional deposition of a thin Au layer onto the entire wafer may be introduced before source deposition to prevent charging of the wafer during the ion-implantation process\cite{Niemeyer2024}. It is worth mentioning that the deposition of the radioactive source is optional. It can be easily omitted in order to, for example, fabricate absorbers with different thicknesses on the same wafer, as mentioned above.

For adding the second absorber half, a single positive-tone photoresist layer is spun onto the wafer and structured such that molds on top of the first absorber half are created during resist development (see Fig.~\ref{fig:FabProcess}j). The second absorber half is then deposited via electroplating (see Fig.~\ref{fig:FabProcess}k). This process step is performed without the use of a dedicated seed layer. Instead, the electrical contact to the cathode of the electroplating setup is realized via the first absorber half, the underlying temperature sensor layer and its electrical connection to the on-chip heat bath realized by a metal layer. After electroplating, the resist is fully removed by immersing the substrate in an organic solvent (see Fig.~\ref{fig:FabProcess}l). The deposition of the second absorber half requires a resist thickness that is larger than the total absorber thickness to ensure that even at the very edge of the first absorber half, the resist thickness exceeds the target thickness of the second absorber half. Here, the topology of the wafer after depositing the first absorber half as well as the viscosity of the photoresist layer required to achieve large resist layer thicknesses are causing an inhomogeneously thick photoresist layer. Moreover, it is worth mentioning that for applying and patterning the thick photoresist layer, the process parameters are not constrained as no underlying resist is existent.

In section~\ref{sec:resist} we discuss in detail the photoresist processes needed for the above described fabrication steps and how they were optimized. The custom electroplating setup developed for this work is described in section~\ref{sec:plating} as well as the optimization of the electroplating parameters. The releasing steps after the deposition of every absorber layer are described in section~\ref{sec:release}. Finally, in section~\ref{sec:demo}, we demonstrate the successful microfabrication of absorbers with $4\pi$ detection geometry and their integration into actual MMC-based detector systems.


\section{Optimization of the photoresist processes}\label{sec:resist}
As discussed above, two main photoresist processes are required for the fabrication of free-standing particle absorbers with $4\pi$ detection geometry: 1) A double-resist stack with seed layer in between for the fabrication of the first absorber half and 2) a single, thick resist for electroplating of the second absorber half. We note that in the course of the project PrimA-LTD, processing of the resist mask used for ion-implantation \cite{Mueller2024PrimALTD, Niemeyer2024} is equal to the second process and that individual chips rather than an entire wafer were processed. All photoresists and developers used in this work were produced by Merck KGaA, Darmstadt, Germany and distributed by MicroChemicals GmbH, Ulm, Germany. For the exposure of all photoresists, we used a laser lithography system / maskless aligner for exposing the photoresist..

We chose the positive-tone photoresist AZ 4533 for the lower resist layer of the double-resist stack, that, spun on at \SI{1900}{rpm} on a \SI{2}{inch} wafer and softbaked at \SI{100}{\celsius} for \SI{3}{min}, provides a thickness of \SI{5}{\micro\m}. Several dose tests resulted in an optimal exposure dose of \SI{360}{mJ/cm^{2}} with a development time of \SI{4}{min} using the developer AZ 2026 MIF. The baking step for thermal reflow, taking place above the softening point of the resist, was tested at \SI{120}{\celsius} for \SI{60}{s} and \SI{120}{s} and at \SI{130}{\celsius} for \SI{120}{s}. Profilometer measurements showed an insignificant increase of the reflow strength with higher baking temperature and duration, thus we fixed the reflow step at \SI{120}{\celsius} for \SI{60}{s}. The deposition of the Au seed layer with a thickness of \SI{80}{nm} on the entire wafer is done by dc-magnetron sputtering with the wafer being attached to a metal sample holder using vacuum grease and without any active cooling.

Before optimizing the upper resist layer, we performed a test series to find the maximum baking temperatures and duration at which the upper resist layer can be processed without disturbance of the underlying structures. For this, test chips with reflowed stem holes and seed layer underwent various baking steps. Baking on a hotplate at \SI{85}{\celsius} for a few minutes already leads to strong wrinkling of the seed layer with increasing baking time due to softening of the lower resist layer. This may lead to an incomplete removal of the seed layer during further process steps. Reducing the hotplate temperature to \SI{80}{\celsius}, baking times can be increased to \SI{45}{min} without issues, whereas weak wrinkling starts at baking times longer than \SI{60}{min}. In order to achieve even higher baking times we found the use of an oven to be much more gentle and baking times of over \SI{16}{h} at \SI{80}{\celsius} could be realized. 

For the upper resist layer of the double-resist stack, we chose the positive photoresist AZ 40XT that can yield thicknesses from \SI{15}{\micro\m} to \SI{100}{\micro\m}. It is optimized specifically for this thickness range and as an chemically amplified resist provides steep edges at reasonably low exposure doses and development times compared to standard positive resists at such thicknesses. Furthermore, no rehydration before exposure is necessary for this resist. The resist is spun on the \SI{2}{inch} wafer with the following spin profile: \SI{1500}{rpm} for \SI{20}{s}, \SI{0}{rpm} for \SI{10}{s} and \SI{7000}{rpm} for \SI{2}{s}. With the additional high-speed step, the width of the edge bead can be significantly reduced. The resulting resist thickness is about \SI{25}{\micro\m}. Taken from the datasheet, the recommended baking temperatures for softbake and post exposure bake (PEB) are specified as \SI{125}{\celsius} and \SI{105}{\celsius}, respectively. However, with the limitation of the baking temperature to \SI{80}{\celsius} to avoid a reflow of the lower resist layer of the double-resist stack, we chose a significantly longer baking time of \SI{16}{h} in a drying oven for compensation. To further minimize the solvent concentration in the vicinity of the substrate surface, we introduced a drying step at room temperature for about \SI{3}{h} before starting the softbake. This greatly improves the adhesion of the resist to the substrate and lowers the dark erosion rate in the substrate-near resist. The exposure is done with a comparably large dose $>\SI{1000}{mJ/cm^{2}}$. We found that, with a single exposure step, photoresist still remains in the holes for the absorber stems after developing, as the resist thickness there is effectively larger (by the thickness of the lower resist layer). Introducing a double-exposure step by additionally exposing the stem areas with a dose of about \SI{300}{mJ/cm^2} enables a sufficiently high development rate of this deep photoresist. In the next step, the post-exposure bake is done on a \SI{80}{\celsius} hotplate for \SI{8}{min}. For resist development, the developer AZ 2026 MIF is used. The structured resist is then hardbaked on a \SI{60}{\celsius} hotplate for \SI{20}{min}.

We observed that very dry resist (in the sense of low water content) tends to soaks up when fully immersed in a heated water bath. This can degrade the structural stability of the photoresist and may also promote contamination of the electrolyte during electroplating. To achieve a very reproducible process, we therefore introduced a hydration step before electroplating. This is done in a sealed container with a humidity of $>\SI{90}{\percent}$ for a duration of $>\SI{12}{h}$. After this hydration step, the wafer is immediately immersed into the electrolyte and the electroplating process as described below is performed.

For the fabrication of the second absorber half, a single photoresist layer of AZ~40XT is processed similarly to above. In the following, we hence mention only differences. After spin coating, room temperature drying of the photoresist is performed for \SI{3}{h}. In this case, it especially prevents the formation of bubbles during the softbake in the vicinity of the first absorber halves, where the resist can get very thick. Due to the absence of a lower photoresist layer, the resist AZ~40XT can be processed at the baking temperatures recommended by the manufacturer. Therefore, the softbake is performed on a \SI{125}{\celsius} proximity hotplate in the following steps: \SI{1.5}{mm} for \SI{5}{min}, \SI{0.5}{mm} for \SI{5}{min} and contact for \SI{3}{min}. The post-exposure bake is done at \SI{105}{\celsius} with \SI{1.5}{mm} for \SI{10}{s}, \SI{0.5}{mm} for \SI{10}{s} and contact for \SI{80}{s}. After development, hardbake and hydration, the second absorber half can be electroplated.

When the fabrication of second absorber halves needs to be performed on separated chips (rather than on a full wafer), e.g. when ion-implantation is chosen as the source preparation technique, spin coating of the AZ~40XT can be performed by using a carrier plate with a recess having the exact chip dimensions. This allows for a better resist flow off the chip and therefore reduces the width of the edge bead. For a separated chip having the dimensions $5\times \SI{25}{mm^{2}}$, as used in the project PrimA-LTD, the spin coating parameters needed to be adjusted to \SI{2200}{rpm} for \SI{20}{s} to yield a photoresist thickness of about \SI{25}{\micro\m}.


\section{Electroplating of highly pure Au layers}\label{sec:plating}

\begin{figure*}
    \centering
    \includegraphics[width=.75\textwidth]{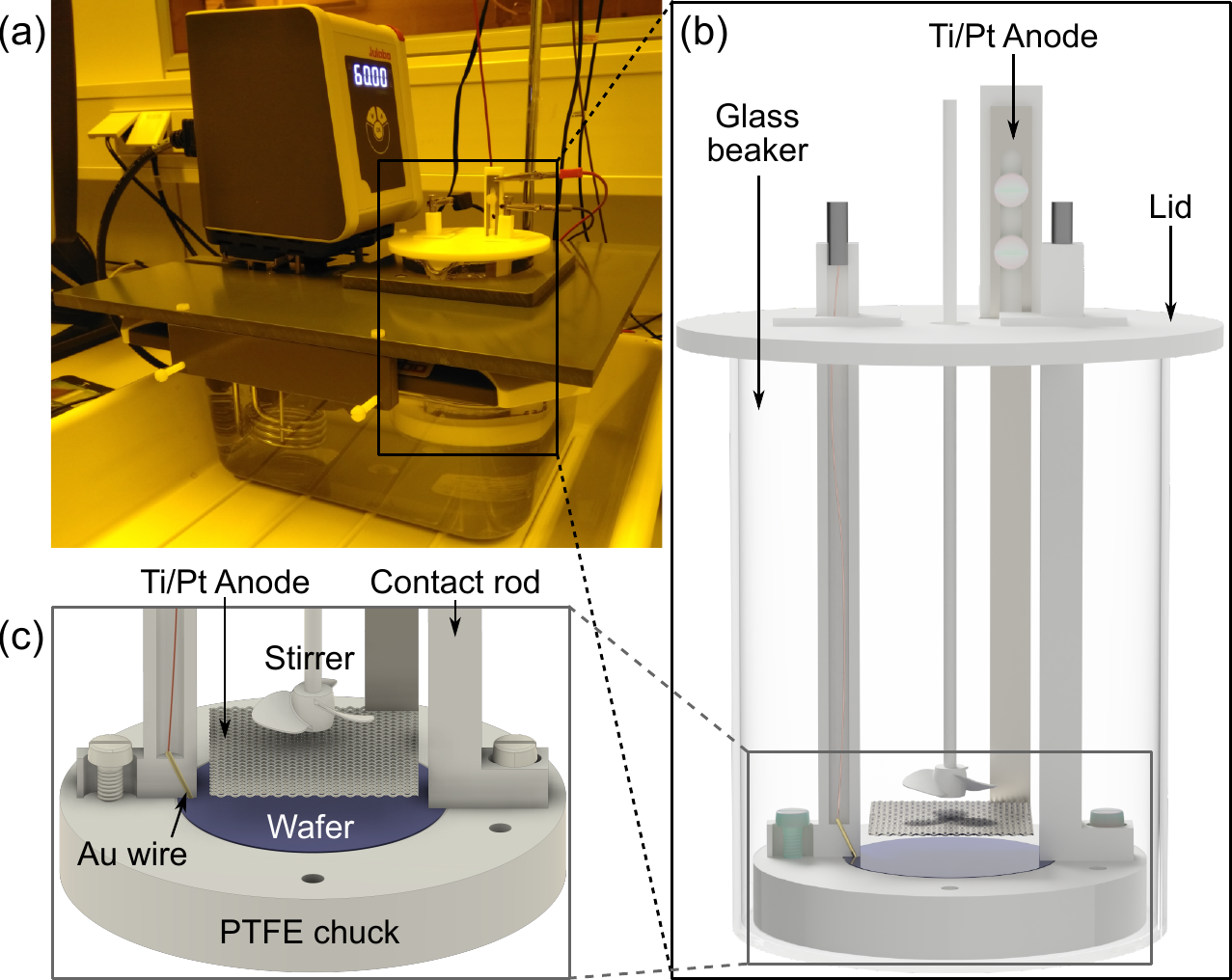}
    \caption{\label{fig:ElectroplatingSetup}Total photograph (a) and 3D render images (b and c) of the setup used for Au electroplating. The glass beaker with electrolyte, anode, contact rods and wafer to be electroplated is mounted into a heated water bath. The beaker is fully covered with a lid and crocodile clamps provide contact to the current source. A stirrer allows creating a bottom-up flow of the electrolyte.}
\end{figure*}

Fig.~\ref{fig:ElectroplatingSetup} shows a total photograph and 3D-rendered close-ups of the electroplating setup developed for this work. To minimize the number of impurities in the deposited Au layer, contamination of the electrolyte has to be prevented during its handling and the actual Au deposition process. For this reason, we chose Polytetrafluoroethylene (PTFE), Polyamide (PA) and DURAN glass (by DWK Life Sciences GmbH, Wertheim/Main, Germany) as non-metallic and inert materials to use in direct contact with the electrolyte. A small sample, e.g. an individual chip, or an entire wafer up to a size of \SI{2}{inch} is mounted onto a custom PTFE plate by contact rods made from PTFE that are screwed to the plate and hold the sample in place. One contact rod is hollow and hosts a copper wire running down inside. This wire is soldered to a short Au cylinder with a diameter of \SI{1}{\milli\m}. The Au cylinder is pressed tightly (to create a liquid-tight sealing) into a small drilling at the bottom of the contact rod, providing a contact point to the sample. With this configuration, we ensure that the copper wire is not in direct contact with the electrolyte to avoid contamination. The setup together with about \SI{200}{\milli\l} of the Au electrolyte TECHNI GOLD 25 ES (by Technic France, Saint-Denis, France) fit into a \SI{1}{\l} DURAN beaker which is covered with a lid that allows the upper ends of the contact rods to pass through but otherwise seals the beaker, preventing the escape of evaporated electrolyte during the process. A custom platinized titanium grid anode supplied by METAKEM (Metakem GmbH, Usingen, Germany) is attached to the lid and its grid is placed in parallel to the surface of the substrate with a distance of about \SI{1}{cm}. The beaker is placed in a temperature-controlled open heating bath circulator CORIO C-BT9 (by JULABO GmbH, Seelbach, Germany). A PTFE stirrer (Mini-Propeller Stirrer by Bohlender GmbH, Grünsfeld, Germany) can be fed through the lid of the beaker to provide a bottom-up axial flow to the electrolyte before or during the electroplating. A Keithley Model 6221 current source is used and connected to both, the anode and the contact rods, by crocodile clamps.

The absolute current that we use for the electroplating process depends on the area to be electroplated. To achieve equal deposition conditions (especially deposition rate), a fixed current density is used across different samples which is defined as the current strength per area to be electroplated. The Au deposition can be performed with either constant-current or pulsed-current electroplating. In the case of constant-current, a continuous current $I_{\mathrm{dc}}$ is applied to the electrolyte over the whole duration of the process. Pulsed-current means, that a current $I_{\mathrm{on}}$ is applied to the electrolyte for a time $t_{\mathrm{on}}$ and withdrawn for $t_{\mathrm{off}}$, periodically. For this, we use the square wave generator of the Keithley Model 6221. The currents used in both process types can be compared by a mean current $I_{\mathrm{ep}} \mathrel{\widehat{=}} I_{\mathrm{on}}[t_{\mathrm{on}}/({t_{\mathrm{on}}+t_{\mathrm{off}}})] \mathrel{\widehat{=}} I_{\mathrm{dc}}$.

For an electroplating process, the sample patterned by a photoresist mask is mounted to the PTFE plate, electrically contacted to dedicated contact points on the wafer by one or two contact rods (depending on the substrate size) and transferred into the beaker. The required amount of electrolyte is filtered through a cellulose nitrate membrane filter by vacuum filtration and is filled into the beaker which is then placed within the heated water bath. During the heating of the electrolyte to the process temperature of $T = \SI{60}{\celsius}$, the electrolyte is stirred for about \SI{20}{\min} with a speed of \SI{600}{rpm} to remove air bubbles from the substrate that form during pouring of the electrolyte. Next, the lid including the anode is placed onto the beaker, the current source is connected and the process is started. No stirring is applied during the electroplating process. After the deposition is finished, the electrolyte is filled back into a glass bottle and the whole setup is thoroughly rinsed with DI water and dried with compressed air.

\begin{figure*}
    \centering
    \includegraphics[width=.8\textwidth]{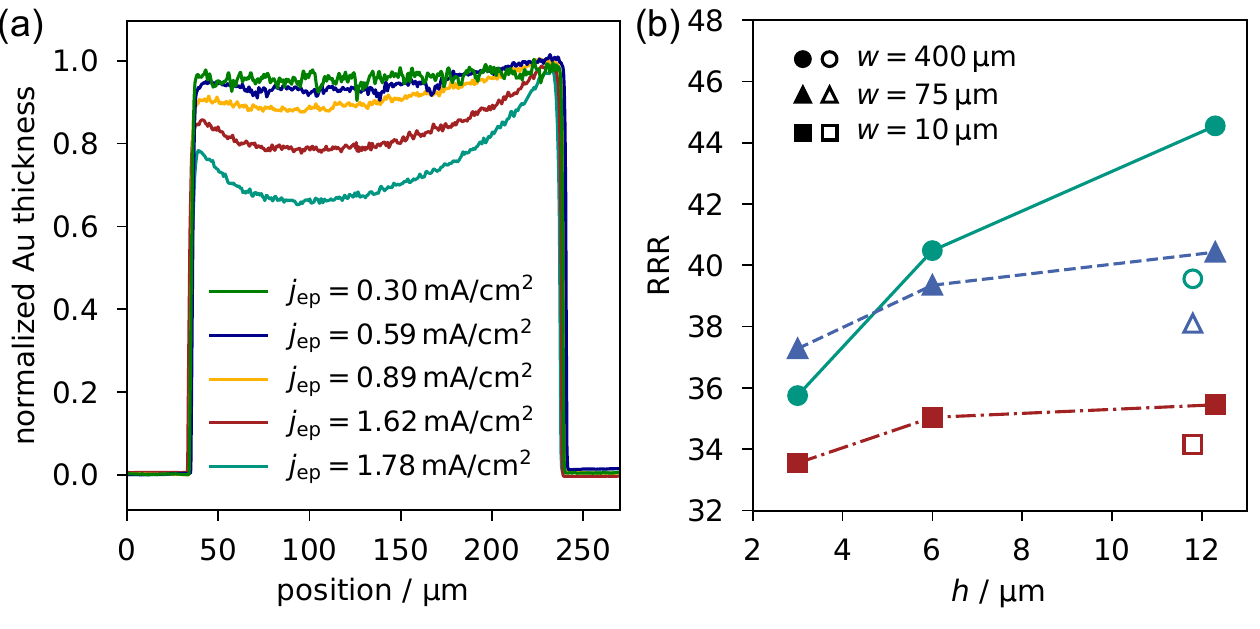}
    \caption{\label{fig:RRR}(a) Measured height profiles of \SI{200}{\micro\m} wide Au stripes electroplated with varying mean current density $j_\mathrm{ep}$. $j_\mathrm{ep}$ was adjusted by variation of $t_\mathrm{off}$ with fixed $j_\mathrm{on} = \SI{1.78}{mA/cm^2}$ and $t_\mathrm{on} = \SI{10}{ms}$. With $j_\mathrm{ep}=\SI{0.30}{mA/cm^{2}}$ excellent thickness homogeneity can be achieved. (b) Measured RRR values of straight Au lines with varying thickness $h$ and linewidth $w$. Samples electroplated with pulsed-current and constant-current are indicated by filled and empty symbols, respectively. Samples with large dimensions achieve $\mathrm{RRR}>44$, while smaller dimensions lead to a slightly reduced RRR. Moreover, Au samples electroplated with pulsed-current yield larger RRR compared to constant-current. All samples were electroplated with a mean current density of $j_\mathrm{ep}=\SI{0.30}{mA/cm^{2}}$.}
\end{figure*}

To achieve good thickness homogeneity of the electroplated absorbers, we performed a test series with \SI{200}{\micro\m} wide Au strips and varying mean current densities $j_\mathrm{ep}$ by variation of the time $t_{\mathrm{off}}$. Fig.~\ref{fig:RRR}a) depicts height profiles of these samples measured with a stylus profiler. We observed, that too large current densities $j_\mathrm{ep}$ lead to a strong thickness inhomogeneity with a characteristic u-shape, that is a result of Au depletion in the substrate-near electrolyte due to insufficient re-diffusion of Au-ions. The observed asymmetry of the u-shape is not fully clarified. It may be induced by deposition of Au-ions in neighboring structures, which leads to an asymmetric depletion of the electrolyte in the measured structure. The detection efficiency of an absorber with such a thickness profile would significantly vary over the absorber area. With a current density $j_\mathrm{ep}=\SI{0.30}{mA/cm^{2}}$, however, sufficient re-diffusion of Au-ions is provided and excellent thickness homogeneity can be achieved. We did not observe a significant difference in thickness homogeneity between constant-current or pulsed-current electroplating.

In Fig.~\ref{fig:RRR}b), measured RRR values of Au test structures are shown. For this, we deposited straight Au lines with different thicknesses $h$ and linewidths $w$ using the photoresist AZ 40XT and the electroplating processes described above. For comparison, we used both, constant-current (empty symbols) and pulsed-current (filled symbols) electroplating, with $j_\mathrm{ep}=\SI{0.30}{mA/cm^{2}}$. In the case of pulsed-current electroplated samples, we used $t_{\mathrm{on}}=\SI{10}{ms}$ and $t_{\mathrm{off}}=\SI{50}{ms}$. The resulting deposition rate was about \SI{17.6}{nm/min} for all samples. A single batch of electrolyte was reused from sample to sample. The data clearly shows that pulsed-current electroplated samples can yield $\mathrm{RRR}>44$, when dimensions are large, i.e. $h=\SI{12}{\micro\m}$ and $w=\SI{400}{\micro\m}$. However, the RRR values of samples with smaller dimensions are slightly reduced down to $\mathrm{RRR}<34$ for $d=\SI{3}{\micro\m}$ and $w=\SI{10}{\micro\m}$ in our test series. The mean free path of the electrons, calculated like shown in \cite{Brown2008}, is in all cases limited by scattering of electrons on film impurities and not by the geometric limitations of the film itself. Nevertheless, the RRR values are in all cases very high, exceeding typical values reported in literature by about a factor of two \cite{Fleischmann2009, Friedrich2016}, and being in the order of the highest values reported in literature \cite{Brown2008, Yagi2023}. Furthermore, we see that pulsed-current electroplated samples show higher RRR than corresponding samples of about the same thickness electroplated by constant-current. We ensured that the changing RRR is not a result of the aging electrolyte batch by a mixed order of sample fabrication. From this test series, we conclude that particle absorbers will have RRR between 40 and 45 with excellent thickness homogeneity, when using pulsed-current electroplating with $t_{\mathrm{on}}=\SI{10}{ms}$, $t_{\mathrm{off}}=\SI{50}{ms}$ and $j_\mathrm{ep}=\SI{0.30}{mA/cm^{2}}$.


\begin{figure*}
    \centering
    \includegraphics[width=.78\textwidth]{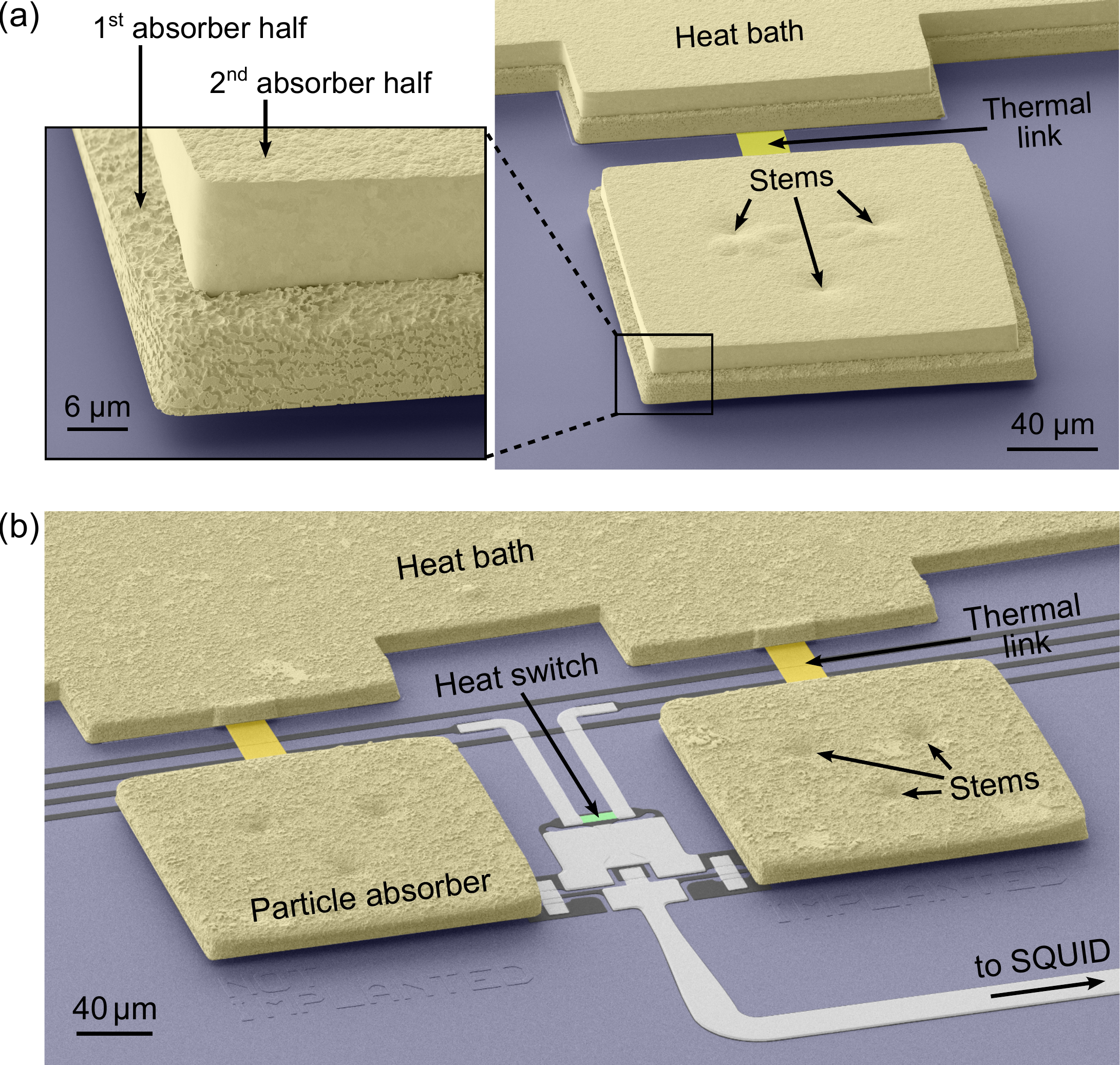}
    \caption{\label{fig:FinalAbsorber}(a) Colorized SEM image of a free standing $4\pi$ absorber composed of separately electroplated first and second absorber halves, having thicknesses of about \SI{11.1}{\micro\m} and \SI{12.6}{\micro\m}, respectively. The roughness of the first absorber half is due to the etching step applied during its fabrication. This demonstration of the fabrication process took place on a dummy wafer without actual MMC on it. (b) Colorized SEM image of two MMC pixels that were fabricated for measuring the decay energy spectrum of $^{55}$Fe in the frame of the PrimA-LTD project. The free-standing absorbers are fabricated up to the first absorber half. Up to this fabrication stage, no $^{55}$Fe is ion-implanted into the absorbers.}
\end{figure*}

\section{Release process for free standing absorber structures}\label{sec:release}
After electroplating of the first absorber half, the release of the absorber is performed by removing the supporting structures layer by layer (see Fig.~\ref{fig:FabProcess}f). For this, the wafer is immediately rinsed with water for several minutes after electroplating to remove any electrolyte residues. Fast drying of the photoresist at ambient humidity leads to cracks in the resist that can peal off large pieces of the underlying seed layer, making the following process steps rather difficult. For this reason, the wafer is immersed in an organic solvent immediately after the rinsing with water in order to remove the upper resist layer. When using Dimethylformamide (DMF), \SI{1}{min} is sufficient for complete removal, without affecting the lower resist layer. When using Acetone, one has to make sure that any electrolyte is completely washed away from the wafer, as strong precipitation can occur otherwise. After drying of the wafer, the Au seed layer is etched using an iodine-potassium iodide solution for up to \SI{15}{min} and the wafer is subsequently rinsed with water and dried. To remove the lower resist layer, the wafer is immersed in Acetone for at least \SI{1}{h}. We observed, that the seed layer can't be etched entirely and a very thin, but electrically conductive, residue is still attached to the absorbers. To remove this residue, the same acetone bath is exposed to a megasonic treatment for \SI{30}{s} (megasonic transducer by SONOSYS Ultrasonic Systems GmbH, Neuenbürg, Germany). Afterwards, the wafer is cleaned in two Isopropanol baths and finally dried. The release after source deposition (Fig.~\ref{fig:FabProcess}i) and electroplating of the second absorber half (Fig.~\ref{fig:FabProcess}l) is much simpler and just performed by immersing the wafer into an organic solvent similarly to above.

\section{Demonstration of the full absorber fabrication process}\label{sec:demo}
For giving a full demonstration of the developed fabrication process, we used a dummy \SI{2}{inch} wafer with a \SI{100}{nm} thick sputtered Au layer structured to imitate the temperature sensor and the heat bath of an MMC. The typical yield with the presented process between \SI{80}{\percent} and \SI{90}{\percent} of absorbers that meet the requirements is very high. Fig.~\ref{fig:FinalAbsorber}a) depicts a colorized scanning electron microscope (SEM) image of one of the fabricated absorbers consisting of a lower and an upper absorber half having thicknesses of about \SI{11.1}{\micro\m} and \SI{12.6}{\micro\m}, respectively. The surface roughness of the first absorber half is a result of the Au wet etching step performed to remove the seed layer (as described above). Such an absorber stack was used within the EMPIR project PrimA-LTD for measuring the electron capture decay spectrum of $^{55}$Fe. Fig.~\ref{fig:FinalAbsorber}b) shows a colorized SEM picture of one of these MMCs fabricated up to the first absorber half (corresponds to Fig.~\ref{fig:FabProcess}f). Both absorbers have a thickness of \SI{11.4}{\micro\m} and are free-standing on three stems each, as can be seen from the shadow underneath the absorbers. As the electron-beam of the SEM may damage the detectors, no image of a completed MMC with fully enclosed $^{55}$Fe source is shown here. For the fabrication on any other detector, the layer thicknesses of both absorber halves can be adjusted as desired by changing the total electroplating time. The thickness of the absorber layers is only limited by the thicknesses of the respective photoresist layers.

\section{Conclusion}
We have successfully developed a microfabrication process for realizing microstructured particle absorbers for cryogenic detectors. The absorbers are composed of two separately electroplated Au layers with thicknesses up to \SI{12}{\micro\m} each. For this, we optimized the necessary photoresist processing and electroplating parameters for the deposition of highly pure Au having residual resistivity ratios above 40. In this way, position-dependencies of the detector signals can be effectively suppressed. Furthermore, the absorbers are free-standing and the microfabrication process is compatible with the embedding of radioactive sources in a $4\pi$ geometry by drop-deposition or ion-implantation, for applications in e.g. radionuclide metrology that require almost unity detection efficiency. Excellent thickness homogeneity of the deposited Au layers ensures a uniform detection efficiency over the entire absorber area. The presented advances in particle absorber fabrication enable new possibilities in a variety of microcalorimeter applications like decay energy spectrometry, where integrated sources need be to measured with excellent energy resolution.

\begin{acknowledgments}
The project 20FUN04 PrimA-LTD has received funding from the EMPIR programme co-financed by the Participating States and from the European Union’s Horizon 2020 research and innovation programme. In addition, M. Müller gratefully acknowledges support by the Karlsruhe School of Elementary Particle and Astroparticle Physics: Science and Technology (KSETA).
\end{acknowledgments}

\section*{Author Declarations}

\subsection*{Conflict of Interest}
The authors have no conflicts to disclose.

\subsection*{Data Availability Statement}
The data that support the findings of this study are available from the corresponding author upon reasonable request.

\subsection*{Author Contributions}
\textbf{Michael Müller:} Conceptualization (equal); Formal analysis (equal); Investigation (equal); Visualization (lead); Writing/Original Draft Preparation (equal). \textbf{Ria-Helen Zühlke:} Formal analysis (supporting); Investigation (equal). \textbf{Sebastian Kempf:} Conceptualization (equal); Formal analysis (equal); Funding acquisition (equal); Investigation (supporting); Project Administration (lead); Supervision (lead) Visualization (supporting); Writing/Original Draft Preparation (equal).

\bibliography{bibliography}

\begin{thebibliography}{30}%
\makeatletter
\providecommand \@ifxundefined [1]{%
 \@ifx{#1\undefined}
}%
\providecommand \@ifnum [1]{%
 \ifnum #1\expandafter \@firstoftwo
 \else \expandafter \@secondoftwo
 \fi
}%
\providecommand \@ifx [1]{%
 \ifx #1\expandafter \@firstoftwo
 \else \expandafter \@secondoftwo
 \fi
}%
\providecommand \natexlab [1]{#1}%
\providecommand \enquote  [1]{``#1''}%
\providecommand \bibnamefont  [1]{#1}%
\providecommand \bibfnamefont [1]{#1}%
\providecommand \citenamefont [1]{#1}%
\providecommand \href@noop [0]{\@secondoftwo}%
\providecommand \href [0]{\begingroup \@sanitize@url \@href}%
\providecommand \@href[1]{\@@startlink{#1}\@@href}%
\providecommand \@@href[1]{\endgroup#1\@@endlink}%
\providecommand \@sanitize@url [0]{\catcode `\\12\catcode `\$12\catcode `\&12\catcode `\#12\catcode `\^12\catcode `\_12\catcode `\%12\relax}%
\providecommand \@@startlink[1]{}%
\providecommand \@@endlink[0]{}%
\providecommand \url  [0]{\begingroup\@sanitize@url \@url }%
\providecommand \@url [1]{\endgroup\@href {#1}{\urlprefix }}%
\providecommand \urlprefix  [0]{URL }%
\providecommand \Eprint [0]{\href }%
\providecommand \doibase [0]{http://dx.doi.org/}%
\providecommand \selectlanguage [0]{\@gobble}%
\providecommand \bibinfo  [0]{\@secondoftwo}%
\providecommand \bibfield  [0]{\@secondoftwo}%
\providecommand \translation [1]{[#1]}%
\providecommand \BibitemOpen [0]{}%
\providecommand \bibitemStop [0]{}%
\providecommand \bibitemNoStop [0]{.\EOS\space}%
\providecommand \EOS [0]{\spacefactor3000\relax}%
\providecommand \BibitemShut  [1]{\csname bibitem#1\endcsname}%
\let\auto@bib@innerbib\@empty
\bibitem [{\citenamefont {Fleischmann}, \citenamefont {Enss},\ and\ \citenamefont {Seidel}(2005)}]{Fleischmann2005}%
  \BibitemOpen
  \bibfield  {author} {\bibinfo {author} {\bibfnamefont {A.}~\bibnamefont {Fleischmann}}, \bibinfo {author} {\bibfnamefont {C.}~\bibnamefont {Enss}}, \ and\ \bibinfo {author} {\bibfnamefont {G.}~\bibnamefont {Seidel}},\ }\enquote {\bibinfo {title} {Metallic magnetic calorimeters},}\ in\ \href {\doibase 10.1007/10933596_4} {\emph {\bibinfo {booktitle} {Cryogenic Particle Detection}}},\ \bibinfo {editor} {edited by\ \bibinfo {editor} {\bibfnamefont {C.}~\bibnamefont {Enss}}}\ (\bibinfo  {publisher} {Springer Berlin Heidelberg},\ \bibinfo {address} {Berlin, Heidelberg},\ \bibinfo {year} {2005})\ pp.\ \bibinfo {pages} {151--216}\BibitemShut {NoStop}%
\bibitem [{\citenamefont {Kempf}\ \emph {et~al.}(2018)\citenamefont {Kempf}, \citenamefont {Fleischmann}, \citenamefont {Gastaldo},\ and\ \citenamefont {Enss}}]{Kempf2018}%
  \BibitemOpen
  \bibfield  {author} {\bibinfo {author} {\bibfnamefont {S.}~\bibnamefont {Kempf}}, \bibinfo {author} {\bibfnamefont {A.}~\bibnamefont {Fleischmann}}, \bibinfo {author} {\bibfnamefont {L.}~\bibnamefont {Gastaldo}}, \ and\ \bibinfo {author} {\bibfnamefont {C.}~\bibnamefont {Enss}},\ }\bibfield  {title} {\enquote {\bibinfo {title} {Physics and applications of metallic magnetic calorimeters},}\ }\href {\doibase 10.1007/s10909-018-1891-6} {\bibfield  {journal} {\bibinfo  {journal} {Journal of Low Temperature Physics}\ }\textbf {\bibinfo {volume} {193}},\ \bibinfo {pages} {365--379} (\bibinfo {year} {2018})}\BibitemShut {NoStop}%
\bibitem [{\citenamefont {Krantz}\ \emph {et~al.}(2024)\citenamefont {Krantz}, \citenamefont {Toschi}, \citenamefont {Maier}, \citenamefont {Heine}, \citenamefont {Enss},\ and\ \citenamefont {Kempf}}]{Krantz2024}%
  \BibitemOpen
  \bibfield  {author} {\bibinfo {author} {\bibfnamefont {M.}~\bibnamefont {Krantz}}, \bibinfo {author} {\bibfnamefont {F.}~\bibnamefont {Toschi}}, \bibinfo {author} {\bibfnamefont {B.}~\bibnamefont {Maier}}, \bibinfo {author} {\bibfnamefont {G.}~\bibnamefont {Heine}}, \bibinfo {author} {\bibfnamefont {C.}~\bibnamefont {Enss}}, \ and\ \bibinfo {author} {\bibfnamefont {S.}~\bibnamefont {Kempf}},\ }\bibfield  {title} {\enquote {\bibinfo {title} {{Magnetic microcalorimeter with paramagnetic temperature sensors and integrated dc-SQUID readout for high-resolution x-ray emission spectroscopy}},}\ }\href {\doibase 10.1063/5.0180903} {\bibfield  {journal} {\bibinfo  {journal} {Applied Physics Letters}\ }\textbf {\bibinfo {volume} {124}},\ \bibinfo {pages} {032601} (\bibinfo {year} {2024})}\BibitemShut {NoStop}%
\bibitem [{\citenamefont {Rotzinger}\ \emph {et~al.}(2008)\citenamefont {Rotzinger}, \citenamefont {Linck}, \citenamefont {Burck}, \citenamefont {Rodrigues}, \citenamefont {Loidl}, \citenamefont {Leblanc}, \citenamefont {Fleischmann}, \citenamefont {Fleischmann},\ and\ \citenamefont {Enss}}]{Rotzinger2008}%
  \BibitemOpen
  \bibfield  {author} {\bibinfo {author} {\bibfnamefont {H.}~\bibnamefont {Rotzinger}}, \bibinfo {author} {\bibfnamefont {M.}~\bibnamefont {Linck}}, \bibinfo {author} {\bibfnamefont {A.}~\bibnamefont {Burck}}, \bibinfo {author} {\bibfnamefont {M.}~\bibnamefont {Rodrigues}}, \bibinfo {author} {\bibfnamefont {M.}~\bibnamefont {Loidl}}, \bibinfo {author} {\bibfnamefont {E.}~\bibnamefont {Leblanc}}, \bibinfo {author} {\bibfnamefont {L.}~\bibnamefont {Fleischmann}}, \bibinfo {author} {\bibfnamefont {A.}~\bibnamefont {Fleischmann}}, \ and\ \bibinfo {author} {\bibfnamefont {C.}~\bibnamefont {Enss}},\ }\bibfield  {title} {\enquote {\bibinfo {title} {Beta spectrometry with magnetic calorimeters},}\ }\href {\doibase 10.1007/s10909-008-9787-5} {\bibfield  {journal} {\bibinfo  {journal} {Journal of Low Temperature Physics}\ }\textbf {\bibinfo {volume} {151}},\ \bibinfo {pages} {1087--1093} (\bibinfo {year} {2008})}\BibitemShut {NoStop}%
\bibitem [{\citenamefont {von Krosigk}\ \emph {et~al.}(2023)\citenamefont {von Krosigk}, \citenamefont {Eitel}, \citenamefont {Enss}, \citenamefont {Ferber}, \citenamefont {Gastaldo}, \citenamefont {Kahlhoefer}, \citenamefont {Kempf}, \citenamefont {Klute}, \citenamefont {Lindemann}, \citenamefont {Schumann}, \citenamefont {Toschi},\ and\ \citenamefont {Valerius}}]{Krosigk2023}%
  \BibitemOpen
  \bibfield  {author} {\bibinfo {author} {\bibfnamefont {B.}~\bibnamefont {von Krosigk}}, \bibinfo {author} {\bibfnamefont {K.}~\bibnamefont {Eitel}}, \bibinfo {author} {\bibfnamefont {C.}~\bibnamefont {Enss}}, \bibinfo {author} {\bibfnamefont {T.}~\bibnamefont {Ferber}}, \bibinfo {author} {\bibfnamefont {L.}~\bibnamefont {Gastaldo}}, \bibinfo {author} {\bibfnamefont {F.}~\bibnamefont {Kahlhoefer}}, \bibinfo {author} {\bibfnamefont {S.}~\bibnamefont {Kempf}}, \bibinfo {author} {\bibfnamefont {M.}~\bibnamefont {Klute}}, \bibinfo {author} {\bibfnamefont {S.}~\bibnamefont {Lindemann}}, \bibinfo {author} {\bibfnamefont {M.}~\bibnamefont {Schumann}}, \bibinfo {author} {\bibfnamefont {F.}~\bibnamefont {Toschi}}, \ and\ \bibinfo {author} {\bibfnamefont {K.}~\bibnamefont {Valerius}},\ }\bibfield  {title} {\enquote {\bibinfo {title} {{DELight: A Direct search Experiment for Light dark matter with superfluid helium}},}\ }\href {\doibase 10.21468/SciPostPhysProc.12.016} {\bibfield  {journal} {\bibinfo  {journal}
  {SciPost Phys. Proc.}\ ,\ \bibinfo {pages} {016}} (\bibinfo {year} {2023})}\BibitemShut {NoStop}%
\bibitem [{\citenamefont {Gastaldo}\ \emph {et~al.}(2017)\citenamefont {Gastaldo}, \citenamefont {Blaum}, \citenamefont {Chrysalidis}, \citenamefont {Day~Goodacre}, \citenamefont {Domula}, \citenamefont {Door}, \citenamefont {Dorrer}, \citenamefont {D{\"u}llmann}, \citenamefont {Eberhardt}, \citenamefont {Eliseev}, \citenamefont {Enss}, \citenamefont {Faessler}, \citenamefont {Filianin}, \citenamefont {Fleischmann}, \citenamefont {Fonnesu}, \citenamefont {Gamer}, \citenamefont {Haas}, \citenamefont {Hassel}, \citenamefont {Hengstler}, \citenamefont {Jochum}, \citenamefont {Johnston}, \citenamefont {Kebschull}, \citenamefont {Kempf}, \citenamefont {Kieck}, \citenamefont {K{\"o}ster}, \citenamefont {Lahiri}, \citenamefont {Maiti}, \citenamefont {Mantegazzini}, \citenamefont {Marsh}, \citenamefont {Neroutsos}, \citenamefont {Novikov}, \citenamefont {Ranitzsch}, \citenamefont {Rothe}, \citenamefont {Rischka}, \citenamefont {Saenz}, \citenamefont {Sander}, \citenamefont {Schneider}, \citenamefont {Scholl},
  \citenamefont {Sch{\"u}ssler}, \citenamefont {Schweiger}, \citenamefont {Simkovic}, \citenamefont {Stora}, \citenamefont {Sz{\"u}cs}, \citenamefont {T{\"u}rler}, \citenamefont {Veinhard}, \citenamefont {Weber}, \citenamefont {Wegner}, \citenamefont {Wendt},\ and\ \citenamefont {Zuber}}]{Gastaldo2017}%
  \BibitemOpen
  \bibfield  {author} {\bibinfo {author} {\bibfnamefont {L.}~\bibnamefont {Gastaldo}}, \bibinfo {author} {\bibfnamefont {K.}~\bibnamefont {Blaum}}, \bibinfo {author} {\bibfnamefont {K.}~\bibnamefont {Chrysalidis}}, \bibinfo {author} {\bibfnamefont {T.}~\bibnamefont {Day~Goodacre}}, \bibinfo {author} {\bibfnamefont {A.}~\bibnamefont {Domula}}, \bibinfo {author} {\bibfnamefont {M.}~\bibnamefont {Door}}, \bibinfo {author} {\bibfnamefont {H.}~\bibnamefont {Dorrer}}, \bibinfo {author} {\bibfnamefont {C.~E.}\ \bibnamefont {D{\"u}llmann}}, \bibinfo {author} {\bibfnamefont {K.}~\bibnamefont {Eberhardt}}, \bibinfo {author} {\bibfnamefont {S.}~\bibnamefont {Eliseev}}, \bibinfo {author} {\bibfnamefont {C.}~\bibnamefont {Enss}}, \bibinfo {author} {\bibfnamefont {A.}~\bibnamefont {Faessler}}, \bibinfo {author} {\bibfnamefont {P.}~\bibnamefont {Filianin}}, \bibinfo {author} {\bibfnamefont {A.}~\bibnamefont {Fleischmann}}, \bibinfo {author} {\bibfnamefont {D.}~\bibnamefont {Fonnesu}}, \bibinfo {author} {\bibfnamefont
  {L.}~\bibnamefont {Gamer}}, \bibinfo {author} {\bibfnamefont {R.}~\bibnamefont {Haas}}, \bibinfo {author} {\bibfnamefont {C.}~\bibnamefont {Hassel}}, \bibinfo {author} {\bibfnamefont {D.}~\bibnamefont {Hengstler}}, \bibinfo {author} {\bibfnamefont {J.}~\bibnamefont {Jochum}}, \bibinfo {author} {\bibfnamefont {K.}~\bibnamefont {Johnston}}, \bibinfo {author} {\bibfnamefont {U.}~\bibnamefont {Kebschull}}, \bibinfo {author} {\bibfnamefont {S.}~\bibnamefont {Kempf}}, \bibinfo {author} {\bibfnamefont {T.}~\bibnamefont {Kieck}}, \bibinfo {author} {\bibfnamefont {U.}~\bibnamefont {K{\"o}ster}}, \bibinfo {author} {\bibfnamefont {S.}~\bibnamefont {Lahiri}}, \bibinfo {author} {\bibfnamefont {M.}~\bibnamefont {Maiti}}, \bibinfo {author} {\bibfnamefont {F.}~\bibnamefont {Mantegazzini}}, \bibinfo {author} {\bibfnamefont {B.}~\bibnamefont {Marsh}}, \bibinfo {author} {\bibfnamefont {P.}~\bibnamefont {Neroutsos}}, \bibinfo {author} {\bibfnamefont {Y.~N.}\ \bibnamefont {Novikov}}, \bibinfo {author} {\bibfnamefont {P.~C.~O.}\
  \bibnamefont {Ranitzsch}}, \bibinfo {author} {\bibfnamefont {S.}~\bibnamefont {Rothe}}, \bibinfo {author} {\bibfnamefont {A.}~\bibnamefont {Rischka}}, \bibinfo {author} {\bibfnamefont {A.}~\bibnamefont {Saenz}}, \bibinfo {author} {\bibfnamefont {O.}~\bibnamefont {Sander}}, \bibinfo {author} {\bibfnamefont {F.}~\bibnamefont {Schneider}}, \bibinfo {author} {\bibfnamefont {S.}~\bibnamefont {Scholl}}, \bibinfo {author} {\bibfnamefont {R.~X.}\ \bibnamefont {Sch{\"u}ssler}}, \bibinfo {author} {\bibfnamefont {C.}~\bibnamefont {Schweiger}}, \bibinfo {author} {\bibfnamefont {F.}~\bibnamefont {Simkovic}}, \bibinfo {author} {\bibfnamefont {T.}~\bibnamefont {Stora}}, \bibinfo {author} {\bibfnamefont {Z.}~\bibnamefont {Sz{\"u}cs}}, \bibinfo {author} {\bibfnamefont {A.}~\bibnamefont {T{\"u}rler}}, \bibinfo {author} {\bibfnamefont {M.}~\bibnamefont {Veinhard}}, \bibinfo {author} {\bibfnamefont {M.}~\bibnamefont {Weber}}, \bibinfo {author} {\bibfnamefont {M.}~\bibnamefont {Wegner}}, \bibinfo {author} {\bibfnamefont
  {K.}~\bibnamefont {Wendt}}, \ and\ \bibinfo {author} {\bibfnamefont {K.}~\bibnamefont {Zuber}},\ }\bibfield  {title} {\enquote {\bibinfo {title} {The electron capture in 163ho experiment -- echo},}\ }\href {\doibase 10.1140/epjst/e2017-70071-y} {\bibfield  {journal} {\bibinfo  {journal} {The European Physical Journal Special Topics}\ }\textbf {\bibinfo {volume} {226}},\ \bibinfo {pages} {1623--1694} (\bibinfo {year} {2017})}\BibitemShut {NoStop}%
\bibitem [{\citenamefont {Sikorsky}\ \emph {et~al.}(2020)\citenamefont {Sikorsky}, \citenamefont {Geist}, \citenamefont {Hengstler}, \citenamefont {Kempf}, \citenamefont {Gastaldo}, \citenamefont {Enss}, \citenamefont {Mokry}, \citenamefont {Runke}, \citenamefont {D\"ullmann}, \citenamefont {Wobrauschek}, \citenamefont {Beeks}, \citenamefont {Rosecker}, \citenamefont {Sterba}, \citenamefont {Kazakov}, \citenamefont {Schumm},\ and\ \citenamefont {Fleischmann}}]{Sikorsky2020}%
  \BibitemOpen
  \bibfield  {author} {\bibinfo {author} {\bibfnamefont {T.}~\bibnamefont {Sikorsky}}, \bibinfo {author} {\bibfnamefont {J.}~\bibnamefont {Geist}}, \bibinfo {author} {\bibfnamefont {D.}~\bibnamefont {Hengstler}}, \bibinfo {author} {\bibfnamefont {S.}~\bibnamefont {Kempf}}, \bibinfo {author} {\bibfnamefont {L.}~\bibnamefont {Gastaldo}}, \bibinfo {author} {\bibfnamefont {C.}~\bibnamefont {Enss}}, \bibinfo {author} {\bibfnamefont {C.}~\bibnamefont {Mokry}}, \bibinfo {author} {\bibfnamefont {J.}~\bibnamefont {Runke}}, \bibinfo {author} {\bibfnamefont {C.~E.}\ \bibnamefont {D\"ullmann}}, \bibinfo {author} {\bibfnamefont {P.}~\bibnamefont {Wobrauschek}}, \bibinfo {author} {\bibfnamefont {K.}~\bibnamefont {Beeks}}, \bibinfo {author} {\bibfnamefont {V.}~\bibnamefont {Rosecker}}, \bibinfo {author} {\bibfnamefont {J.~H.}\ \bibnamefont {Sterba}}, \bibinfo {author} {\bibfnamefont {G.}~\bibnamefont {Kazakov}}, \bibinfo {author} {\bibfnamefont {T.}~\bibnamefont {Schumm}}, \ and\ \bibinfo {author} {\bibfnamefont
  {A.}~\bibnamefont {Fleischmann}},\ }\bibfield  {title} {\enquote {\bibinfo {title} {Measurement of the $^{229}\mathrm{Th}$ isomer energy with a magnetic microcalorimeter},}\ }\href {\doibase 10.1103/PhysRevLett.125.142503} {\bibfield  {journal} {\bibinfo  {journal} {Phys. Rev. Lett.}\ }\textbf {\bibinfo {volume} {125}},\ \bibinfo {pages} {142503} (\bibinfo {year} {2020})}\BibitemShut {NoStop}%
\bibitem [{\citenamefont {Hengstler}\ \emph {et~al.}(2015)\citenamefont {Hengstler}, \citenamefont {Keller}, \citenamefont {Schötz}, \citenamefont {Geist}, \citenamefont {Krantz}, \citenamefont {Kempf}, \citenamefont {Gastaldo}, \citenamefont {Fleischmann}, \citenamefont {Gassner}, \citenamefont {Weber}, \citenamefont {Märtin}, \citenamefont {Stöhlker},\ and\ \citenamefont {Enss}}]{Hengstler2015}%
  \BibitemOpen
  \bibfield  {author} {\bibinfo {author} {\bibfnamefont {D.}~\bibnamefont {Hengstler}}, \bibinfo {author} {\bibfnamefont {M.}~\bibnamefont {Keller}}, \bibinfo {author} {\bibfnamefont {C.}~\bibnamefont {Schötz}}, \bibinfo {author} {\bibfnamefont {J.}~\bibnamefont {Geist}}, \bibinfo {author} {\bibfnamefont {M.}~\bibnamefont {Krantz}}, \bibinfo {author} {\bibfnamefont {S.}~\bibnamefont {Kempf}}, \bibinfo {author} {\bibfnamefont {L.}~\bibnamefont {Gastaldo}}, \bibinfo {author} {\bibfnamefont {A.}~\bibnamefont {Fleischmann}}, \bibinfo {author} {\bibfnamefont {T.}~\bibnamefont {Gassner}}, \bibinfo {author} {\bibfnamefont {G.}~\bibnamefont {Weber}}, \bibinfo {author} {\bibfnamefont {R.}~\bibnamefont {Märtin}}, \bibinfo {author} {\bibfnamefont {T.}~\bibnamefont {Stöhlker}}, \ and\ \bibinfo {author} {\bibfnamefont {C.}~\bibnamefont {Enss}},\ }\bibfield  {title} {\enquote {\bibinfo {title} {Towards fair: first measurements of metallic magnetic calorimeters for high-resolution x-ray spectroscopy at gsi},}\ }\href
  {\doibase 10.1088/0031-8949/2015/T166/014054} {\bibfield  {journal} {\bibinfo  {journal} {Physica Scripta}\ }\textbf {\bibinfo {volume} {2015}},\ \bibinfo {pages} {014054} (\bibinfo {year} {2015})}\BibitemShut {NoStop}%
\bibitem [{\citenamefont {Pfäfflein}\ \emph {et~al.}(2022)\citenamefont {Pfäfflein}, \citenamefont {Allgeier}, \citenamefont {Bernitt}, \citenamefont {Fleischmann}, \citenamefont {Friedrich}, \citenamefont {Hahn}, \citenamefont {Hengstler}, \citenamefont {Herdrich}, \citenamefont {Kalinin}, \citenamefont {Kröger}, \citenamefont {Kuntz}, \citenamefont {Lestinsky}, \citenamefont {Löher}, \citenamefont {Menz}, \citenamefont {Over}, \citenamefont {Spillmann}, \citenamefont {Weber}, \citenamefont {Zhu}, \citenamefont {Enss},\ and\ \citenamefont {Stöhlker}}]{Pfaefflein2022}%
  \BibitemOpen
  \bibfield  {author} {\bibinfo {author} {\bibfnamefont {P.}~\bibnamefont {Pfäfflein}}, \bibinfo {author} {\bibfnamefont {S.}~\bibnamefont {Allgeier}}, \bibinfo {author} {\bibfnamefont {S.}~\bibnamefont {Bernitt}}, \bibinfo {author} {\bibfnamefont {A.}~\bibnamefont {Fleischmann}}, \bibinfo {author} {\bibfnamefont {M.}~\bibnamefont {Friedrich}}, \bibinfo {author} {\bibfnamefont {C.}~\bibnamefont {Hahn}}, \bibinfo {author} {\bibfnamefont {D.}~\bibnamefont {Hengstler}}, \bibinfo {author} {\bibfnamefont {M.~O.}\ \bibnamefont {Herdrich}}, \bibinfo {author} {\bibfnamefont {A.}~\bibnamefont {Kalinin}}, \bibinfo {author} {\bibfnamefont {F.~M.}\ \bibnamefont {Kröger}}, \bibinfo {author} {\bibfnamefont {P.}~\bibnamefont {Kuntz}}, \bibinfo {author} {\bibfnamefont {M.}~\bibnamefont {Lestinsky}}, \bibinfo {author} {\bibfnamefont {B.}~\bibnamefont {Löher}}, \bibinfo {author} {\bibfnamefont {E.~B.}\ \bibnamefont {Menz}}, \bibinfo {author} {\bibfnamefont {T.}~\bibnamefont {Over}}, \bibinfo {author} {\bibfnamefont
  {U.}~\bibnamefont {Spillmann}}, \bibinfo {author} {\bibfnamefont {G.}~\bibnamefont {Weber}}, \bibinfo {author} {\bibfnamefont {B.}~\bibnamefont {Zhu}}, \bibinfo {author} {\bibfnamefont {C.}~\bibnamefont {Enss}}, \ and\ \bibinfo {author} {\bibfnamefont {T.}~\bibnamefont {Stöhlker}},\ }\bibfield  {title} {\enquote {\bibinfo {title} {Integration of maxs-type microcalorimeter detectors for high-resolution x-ray spectroscopy into the experimental environment at the cryring@esr electron cooler},}\ }\href {\doibase 10.1088/1402-4896/ac93be} {\bibfield  {journal} {\bibinfo  {journal} {Physica Scripta}\ }\textbf {\bibinfo {volume} {97}},\ \bibinfo {pages} {114005} (\bibinfo {year} {2022})}\BibitemShut {NoStop}%
\bibitem [{\citenamefont {Herdrich}\ \emph {et~al.}(2024)\citenamefont {Herdrich}, \citenamefont {Hengstler}, \citenamefont {Allgeier}, \citenamefont {Friedrich}, \citenamefont {Fleischmann}, \citenamefont {Enss}, \citenamefont {Bernitt}, \citenamefont {Morgenroth}, \citenamefont {Trotsenko}, \citenamefont {Schuch},\ and\ \citenamefont {Stöhlker}}]{Herdrich2024}%
  \BibitemOpen
  \bibfield  {author} {\bibinfo {author} {\bibfnamefont {M.~O.}\ \bibnamefont {Herdrich}}, \bibinfo {author} {\bibfnamefont {D.}~\bibnamefont {Hengstler}}, \bibinfo {author} {\bibfnamefont {S.}~\bibnamefont {Allgeier}}, \bibinfo {author} {\bibfnamefont {M.}~\bibnamefont {Friedrich}}, \bibinfo {author} {\bibfnamefont {A.}~\bibnamefont {Fleischmann}}, \bibinfo {author} {\bibfnamefont {C.}~\bibnamefont {Enss}}, \bibinfo {author} {\bibfnamefont {S.}~\bibnamefont {Bernitt}}, \bibinfo {author} {\bibfnamefont {T.}~\bibnamefont {Morgenroth}}, \bibinfo {author} {\bibfnamefont {S.}~\bibnamefont {Trotsenko}}, \bibinfo {author} {\bibfnamefont {R.}~\bibnamefont {Schuch}}, \ and\ \bibinfo {author} {\bibfnamefont {T.}~\bibnamefont {Stöhlker}},\ }\bibfield  {title} {\enquote {\bibinfo {title} {Application of a metallic-magnetic calorimeter for high-resolution x-ray spectroscopy of fe at an ebit},}\ }\href {\doibase 10.1088/1361-6455/ad34a2} {\bibfield  {journal} {\bibinfo  {journal} {Journal of Physics B: Atomic, Molecular
  and Optical Physics}\ }\textbf {\bibinfo {volume} {57}},\ \bibinfo {pages} {085001} (\bibinfo {year} {2024})}\BibitemShut {NoStop}%
\bibitem [{\citenamefont {Alenkov}\ \emph {et~al.}(2019)\citenamefont {Alenkov}, \citenamefont {Bae}, \citenamefont {Beyer}, \citenamefont {Boiko}, \citenamefont {Boonin}, \citenamefont {Buzanov}, \citenamefont {Chanthima}, \citenamefont {Cheoun}, \citenamefont {Chernyak}, \citenamefont {Choe}, \citenamefont {Choi}, \citenamefont {Danevich}, \citenamefont {Djamal}, \citenamefont {Drung}, \citenamefont {Enss}, \citenamefont {Fleischmann}, \citenamefont {Gangapshev}, \citenamefont {Gastaldo}, \citenamefont {Gavriljuk}, \citenamefont {Gezhaev}, \citenamefont {Grigoryeva}, \citenamefont {Gurentsov}, \citenamefont {Gylova}, \citenamefont {Ha}, \citenamefont {Ha}, \citenamefont {Ha}, \citenamefont {Hahn}, \citenamefont {Jang}, \citenamefont {Jeon}, \citenamefont {Jeon}, \citenamefont {Jo}, \citenamefont {Kaewkhao}, \citenamefont {Kang}, \citenamefont {Kang}, \citenamefont {Kang}, \citenamefont {Kazalov}, \citenamefont {Kempf}, \citenamefont {Khan}, \citenamefont {Khan}, \citenamefont {Kim}, \citenamefont {Kim},
  \citenamefont {Kim}, \citenamefont {Kim}, \citenamefont {Kim}, \citenamefont {Kim}, \citenamefont {Kim}, \citenamefont {Kim}, \citenamefont {Kim}, \citenamefont {Kim}, \citenamefont {Kim}, \citenamefont {Kim}, \citenamefont {Kim}, \citenamefont {Kim}, \citenamefont {Kirdsiri}, \citenamefont {Ko}, \citenamefont {Kobychev}, \citenamefont {Kornoukhov}, \citenamefont {Kuzminov}, \citenamefont {Kwon}, \citenamefont {Lee}, \citenamefont {Lee}, \citenamefont {Lee}, \citenamefont {Lee}, \citenamefont {Lee}, \citenamefont {Lee}, \citenamefont {Lee}, \citenamefont {Lee}, \citenamefont {Lee}, \citenamefont {Lee}, \citenamefont {Lee}, \citenamefont {Leonard}, \citenamefont {Li}, \citenamefont {Li}, \citenamefont {Limkitjaroenporn}, \citenamefont {Makarov}, \citenamefont {Oh}, \citenamefont {Oh}, \citenamefont {Olsen}, \citenamefont {Pabitra}, \citenamefont {Panasenko}, \citenamefont {Pandey}, \citenamefont {Park}, \citenamefont {Park}, \citenamefont {Park}, \citenamefont {Park}, \citenamefont {Park}, \citenamefont
  {Poda}, \citenamefont {Polischuk}, \citenamefont {Prihtiadi}, \citenamefont {Ra}, \citenamefont {Ratkevich}, \citenamefont {Rooh}, \citenamefont {Sari}, \citenamefont {Seo}, \citenamefont {Shin}, \citenamefont {Shin}, \citenamefont {Shlegel}, \citenamefont {Siyeon}, \citenamefont {So}, \citenamefont {Son}, \citenamefont {Srisittipokakun}, \citenamefont {Sujita}, \citenamefont {Tretyak}, \citenamefont {Wirawan}, \citenamefont {Woo}, \citenamefont {Yoon}, \citenamefont {Yue},\ and\ \citenamefont {Zaman}}]{Alenkov2019DoubleBetaDecay}%
  \BibitemOpen
  \bibfield  {author} {\bibinfo {author} {\bibfnamefont {V.}~\bibnamefont {Alenkov}}, \bibinfo {author} {\bibfnamefont {H.~W.}\ \bibnamefont {Bae}}, \bibinfo {author} {\bibfnamefont {J.}~\bibnamefont {Beyer}}, \bibinfo {author} {\bibfnamefont {R.~S.}\ \bibnamefont {Boiko}}, \bibinfo {author} {\bibfnamefont {K.}~\bibnamefont {Boonin}}, \bibinfo {author} {\bibfnamefont {O.}~\bibnamefont {Buzanov}}, \bibinfo {author} {\bibfnamefont {N.}~\bibnamefont {Chanthima}}, \bibinfo {author} {\bibfnamefont {M.~K.}\ \bibnamefont {Cheoun}}, \bibinfo {author} {\bibfnamefont {D.~M.}\ \bibnamefont {Chernyak}}, \bibinfo {author} {\bibfnamefont {J.~S.}\ \bibnamefont {Choe}}, \bibinfo {author} {\bibfnamefont {S.}~\bibnamefont {Choi}}, \bibinfo {author} {\bibfnamefont {F.~A.}\ \bibnamefont {Danevich}}, \bibinfo {author} {\bibfnamefont {M.}~\bibnamefont {Djamal}}, \bibinfo {author} {\bibfnamefont {D.}~\bibnamefont {Drung}}, \bibinfo {author} {\bibfnamefont {C.}~\bibnamefont {Enss}}, \bibinfo {author} {\bibfnamefont {A.}~\bibnamefont
  {Fleischmann}}, \bibinfo {author} {\bibfnamefont {A.~M.}\ \bibnamefont {Gangapshev}}, \bibinfo {author} {\bibfnamefont {L.}~\bibnamefont {Gastaldo}}, \bibinfo {author} {\bibfnamefont {Y.~M.}\ \bibnamefont {Gavriljuk}}, \bibinfo {author} {\bibfnamefont {A.~M.}\ \bibnamefont {Gezhaev}}, \bibinfo {author} {\bibfnamefont {V.~D.}\ \bibnamefont {Grigoryeva}}, \bibinfo {author} {\bibfnamefont {V.~I.}\ \bibnamefont {Gurentsov}}, \bibinfo {author} {\bibfnamefont {O.}~\bibnamefont {Gylova}}, \bibinfo {author} {\bibfnamefont {C.}~\bibnamefont {Ha}}, \bibinfo {author} {\bibfnamefont {D.~H.}\ \bibnamefont {Ha}}, \bibinfo {author} {\bibfnamefont {E.~J.}\ \bibnamefont {Ha}}, \bibinfo {author} {\bibfnamefont {I.~S.}\ \bibnamefont {Hahn}}, \bibinfo {author} {\bibfnamefont {C.~H.}\ \bibnamefont {Jang}}, \bibinfo {author} {\bibfnamefont {E.~J.}\ \bibnamefont {Jeon}}, \bibinfo {author} {\bibfnamefont {J.~A.}\ \bibnamefont {Jeon}}, \bibinfo {author} {\bibfnamefont {H.~S.}\ \bibnamefont {Jo}}, \bibinfo {author} {\bibfnamefont
  {J.}~\bibnamefont {Kaewkhao}}, \bibinfo {author} {\bibfnamefont {C.~S.}\ \bibnamefont {Kang}}, \bibinfo {author} {\bibfnamefont {S.~J.}\ \bibnamefont {Kang}}, \bibinfo {author} {\bibfnamefont {W.~G.}\ \bibnamefont {Kang}}, \bibinfo {author} {\bibfnamefont {V.~V.}\ \bibnamefont {Kazalov}}, \bibinfo {author} {\bibfnamefont {S.}~\bibnamefont {Kempf}}, \bibinfo {author} {\bibfnamefont {A.}~\bibnamefont {Khan}}, \bibinfo {author} {\bibfnamefont {S.}~\bibnamefont {Khan}}, \bibinfo {author} {\bibfnamefont {D.~Y.}\ \bibnamefont {Kim}}, \bibinfo {author} {\bibfnamefont {G.~W.}\ \bibnamefont {Kim}}, \bibinfo {author} {\bibfnamefont {H.~B.}\ \bibnamefont {Kim}}, \bibinfo {author} {\bibfnamefont {H.~J.}\ \bibnamefont {Kim}}, \bibinfo {author} {\bibfnamefont {H.~L.}\ \bibnamefont {Kim}}, \bibinfo {author} {\bibfnamefont {H.~S.}\ \bibnamefont {Kim}}, \bibinfo {author} {\bibfnamefont {I.}~\bibnamefont {Kim}}, \bibinfo {author} {\bibfnamefont {S.~C.}\ \bibnamefont {Kim}}, \bibinfo {author} {\bibfnamefont {S.~G.}\
  \bibnamefont {Kim}}, \bibinfo {author} {\bibfnamefont {S.~K.}\ \bibnamefont {Kim}}, \bibinfo {author} {\bibfnamefont {S.~R.}\ \bibnamefont {Kim}}, \bibinfo {author} {\bibfnamefont {W.~T.}\ \bibnamefont {Kim}}, \bibinfo {author} {\bibfnamefont {Y.~D.}\ \bibnamefont {Kim}}, \bibinfo {author} {\bibfnamefont {Y.~H.}\ \bibnamefont {Kim}}, \bibinfo {author} {\bibfnamefont {K.}~\bibnamefont {Kirdsiri}}, \bibinfo {author} {\bibfnamefont {Y.~J.}\ \bibnamefont {Ko}}, \bibinfo {author} {\bibfnamefont {V.~V.}\ \bibnamefont {Kobychev}}, \bibinfo {author} {\bibfnamefont {V.}~\bibnamefont {Kornoukhov}}, \bibinfo {author} {\bibfnamefont {V.~V.}\ \bibnamefont {Kuzminov}}, \bibinfo {author} {\bibfnamefont {D.~H.}\ \bibnamefont {Kwon}}, \bibinfo {author} {\bibfnamefont {C.}~\bibnamefont {Lee}}, \bibinfo {author} {\bibfnamefont {E.~K.}\ \bibnamefont {Lee}}, \bibinfo {author} {\bibfnamefont {H.~J.}\ \bibnamefont {Lee}}, \bibinfo {author} {\bibfnamefont {H.~S.}\ \bibnamefont {Lee}}, \bibinfo {author} {\bibfnamefont {J.~S.}\
  \bibnamefont {Lee}}, \bibinfo {author} {\bibfnamefont {J.~Y.}\ \bibnamefont {Lee}}, \bibinfo {author} {\bibfnamefont {K.~B.}\ \bibnamefont {Lee}}, \bibinfo {author} {\bibfnamefont {M.~H.}\ \bibnamefont {Lee}}, \bibinfo {author} {\bibfnamefont {M.~K.}\ \bibnamefont {Lee}}, \bibinfo {author} {\bibfnamefont {S.~W.}\ \bibnamefont {Lee}}, \bibinfo {author} {\bibfnamefont {S.~H.}\ \bibnamefont {Lee}}, \bibinfo {author} {\bibfnamefont {D.}~\bibnamefont {Leonard}}, \bibinfo {author} {\bibfnamefont {J.}~\bibnamefont {Li}}, \bibinfo {author} {\bibfnamefont {Y.}~\bibnamefont {Li}}, \bibinfo {author} {\bibfnamefont {P.}~\bibnamefont {Limkitjaroenporn}}, \bibinfo {author} {\bibfnamefont {E.~P.}\ \bibnamefont {Makarov}}, \bibinfo {author} {\bibfnamefont {S.~Y.}\ \bibnamefont {Oh}}, \bibinfo {author} {\bibfnamefont {Y.~M.}\ \bibnamefont {Oh}}, \bibinfo {author} {\bibfnamefont {S.~L.}\ \bibnamefont {Olsen}}, \bibinfo {author} {\bibfnamefont {A.}~\bibnamefont {Pabitra}}, \bibinfo {author} {\bibfnamefont {S.~I.}\
  \bibnamefont {Panasenko}}, \bibinfo {author} {\bibfnamefont {I.}~\bibnamefont {Pandey}}, \bibinfo {author} {\bibfnamefont {C.~W.}\ \bibnamefont {Park}}, \bibinfo {author} {\bibfnamefont {H.~K.}\ \bibnamefont {Park}}, \bibinfo {author} {\bibfnamefont {H.~S.}\ \bibnamefont {Park}}, \bibinfo {author} {\bibfnamefont {K.~S.}\ \bibnamefont {Park}}, \bibinfo {author} {\bibfnamefont {S.~Y.}\ \bibnamefont {Park}}, \bibinfo {author} {\bibfnamefont {D.~V.}\ \bibnamefont {Poda}}, \bibinfo {author} {\bibfnamefont {O.~G.}\ \bibnamefont {Polischuk}}, \bibinfo {author} {\bibfnamefont {H.}~\bibnamefont {Prihtiadi}}, \bibinfo {author} {\bibfnamefont {S.~J.}\ \bibnamefont {Ra}}, \bibinfo {author} {\bibfnamefont {S.~S.}\ \bibnamefont {Ratkevich}}, \bibinfo {author} {\bibfnamefont {G.}~\bibnamefont {Rooh}}, \bibinfo {author} {\bibfnamefont {M.~B.}\ \bibnamefont {Sari}}, \bibinfo {author} {\bibfnamefont {K.~M.}\ \bibnamefont {Seo}}, \bibinfo {author} {\bibfnamefont {J.~W.}\ \bibnamefont {Shin}}, \bibinfo {author} {\bibfnamefont
  {K.~A.}\ \bibnamefont {Shin}}, \bibinfo {author} {\bibfnamefont {V.~N.}\ \bibnamefont {Shlegel}}, \bibinfo {author} {\bibfnamefont {K.}~\bibnamefont {Siyeon}}, \bibinfo {author} {\bibfnamefont {J.~H.}\ \bibnamefont {So}}, \bibinfo {author} {\bibfnamefont {J.~K.}\ \bibnamefont {Son}}, \bibinfo {author} {\bibfnamefont {N.}~\bibnamefont {Srisittipokakun}}, \bibinfo {author} {\bibfnamefont {K.}~\bibnamefont {Sujita}}, \bibinfo {author} {\bibfnamefont {V.~I.}\ \bibnamefont {Tretyak}}, \bibinfo {author} {\bibfnamefont {R.}~\bibnamefont {Wirawan}}, \bibinfo {author} {\bibfnamefont {K.~R.}\ \bibnamefont {Woo}}, \bibinfo {author} {\bibfnamefont {Y.~S.}\ \bibnamefont {Yoon}}, \bibinfo {author} {\bibfnamefont {Q.}~\bibnamefont {Yue}}, \ and\ \bibinfo {author} {\bibfnamefont {S.~U.}\ \bibnamefont {Zaman}},\ }\bibfield  {title} {\enquote {\bibinfo {title} {First results from the amore-pilot neutrinoless double beta decay experiment},}\ }\href {\doibase 10.1140/epjc/s10052-019-7279-1} {\bibfield  {journal} {\bibinfo
  {journal} {The European Physical Journal C}\ }\textbf {\bibinfo {volume} {79}},\ \bibinfo {pages} {791} (\bibinfo {year} {2019})}\BibitemShut {NoStop}%
\bibitem [{\citenamefont {Kim}, \citenamefont {Kim},\ and\ \citenamefont {Woo}(2023)}]{Kim2023DoubleBetaDecay}%
  \BibitemOpen
  \bibfield  {author} {\bibinfo {author} {\bibfnamefont {H.}~\bibnamefont {Kim}}, \bibinfo {author} {\bibfnamefont {Y.-H.}\ \bibnamefont {Kim}}, \ and\ \bibinfo {author} {\bibfnamefont {K.-R.}\ \bibnamefont {Woo}},\ }\bibfield  {title} {\enquote {\bibinfo {title} {Cryogenic particle detection based on magnetic microcalorimeters for rare event searches},}\ }\href {\doibase 10.1140/epjp/s13360-023-04068-0} {\bibfield  {journal} {\bibinfo  {journal} {The European Physical Journal Plus}\ }\textbf {\bibinfo {volume} {138}},\ \bibinfo {pages} {518} (\bibinfo {year} {2023})}\BibitemShut {NoStop}%
\bibitem [{\citenamefont {Stevenson}\ \emph {et~al.}(2019)\citenamefont {Stevenson}, \citenamefont {Balvin}, \citenamefont {Bandler}, \citenamefont {Devasia}, \citenamefont {Nagler}, \citenamefont {Ryu}, \citenamefont {Smith},\ and\ \citenamefont {Yoon}}]{Stevenson2019HydraMMC}%
  \BibitemOpen
  \bibfield  {author} {\bibinfo {author} {\bibfnamefont {T.~R.}\ \bibnamefont {Stevenson}}, \bibinfo {author} {\bibfnamefont {M.~A.}\ \bibnamefont {Balvin}}, \bibinfo {author} {\bibfnamefont {S.~R.}\ \bibnamefont {Bandler}}, \bibinfo {author} {\bibfnamefont {A.~M.}\ \bibnamefont {Devasia}}, \bibinfo {author} {\bibfnamefont {P.~C.}\ \bibnamefont {Nagler}}, \bibinfo {author} {\bibfnamefont {K.~K.}\ \bibnamefont {Ryu}}, \bibinfo {author} {\bibfnamefont {S.~J.}\ \bibnamefont {Smith}}, \ and\ \bibinfo {author} {\bibfnamefont {W.}~\bibnamefont {Yoon}},\ }\bibfield  {title} {\enquote {\bibinfo {title} {{Magnetic calorimeter option for the Lynx x-ray microcalorimeter}},}\ }\href {\doibase 10.1117/1.JATIS.5.2.021009} {\bibfield  {journal} {\bibinfo  {journal} {Journal of Astronomical Telescopes, Instruments, and Systems}\ }\textbf {\bibinfo {volume} {5}},\ \bibinfo {pages} {021009} (\bibinfo {year} {2019})}\BibitemShut {NoStop}%
\bibitem [{\citenamefont {Devasia}\ \emph {et~al.}(2022)\citenamefont {Devasia}, \citenamefont {Bandler}, \citenamefont {Ryu}, \citenamefont {Stevenson},\ and\ \citenamefont {Yoon}}]{Devasia2022LynxMMC}%
  \BibitemOpen
  \bibfield  {author} {\bibinfo {author} {\bibfnamefont {A.~M.}\ \bibnamefont {Devasia}}, \bibinfo {author} {\bibfnamefont {S.~R.}\ \bibnamefont {Bandler}}, \bibinfo {author} {\bibfnamefont {K.}~\bibnamefont {Ryu}}, \bibinfo {author} {\bibfnamefont {T.~R.}\ \bibnamefont {Stevenson}}, \ and\ \bibinfo {author} {\bibfnamefont {W.}~\bibnamefont {Yoon}},\ }\bibfield  {title} {\enquote {\bibinfo {title} {Large-scale magnetic microcalorimeter arrays for the lynx x-ray microcalorimeter},}\ }\href {\doibase 10.1007/s10909-022-02767-z} {\bibfield  {journal} {\bibinfo  {journal} {Journal of Low Temperature Physics}\ }\textbf {\bibinfo {volume} {209}},\ \bibinfo {pages} {337--345} (\bibinfo {year} {2022})}\BibitemShut {NoStop}%
\bibitem [{\citenamefont {Müller}\ \emph {et~al.}(2024)\citenamefont {Müller}, \citenamefont {Rodrigues}, \citenamefont {Beyer}, \citenamefont {Loidl},\ and\ \citenamefont {Kempf}}]{Mueller2024PrimALTD}%
  \BibitemOpen
  \bibfield  {author} {\bibinfo {author} {\bibfnamefont {M.}~\bibnamefont {Müller}}, \bibinfo {author} {\bibfnamefont {M.}~\bibnamefont {Rodrigues}}, \bibinfo {author} {\bibfnamefont {J.}~\bibnamefont {Beyer}}, \bibinfo {author} {\bibfnamefont {M.}~\bibnamefont {Loidl}}, \ and\ \bibinfo {author} {\bibfnamefont {S.}~\bibnamefont {Kempf}},\ }\bibfield  {title} {\enquote {\bibinfo {title} {Magnetic microcalorimeters for primary activity standardization within the empir project prima-ltd},}\ }\href {\doibase 10.1007/s10909-024-03048-7} {\bibfield  {journal} {\bibinfo  {journal} {Journal of Low Temperature Physics}\ }\textbf {\bibinfo {volume} {214}},\ \bibinfo {pages} {263--271} (\bibinfo {year} {2024})}\BibitemShut {NoStop}%
\bibitem [{\citenamefont {Kossert}\ \emph {et~al.}(2022)\citenamefont {Kossert}, \citenamefont {Loidl}, \citenamefont {Mougeot}, \citenamefont {Paulsen}, \citenamefont {Ranitzsch},\ and\ \citenamefont {Rodrigues}}]{Kossert2022}%
  \BibitemOpen
  \bibfield  {author} {\bibinfo {author} {\bibfnamefont {K.}~\bibnamefont {Kossert}}, \bibinfo {author} {\bibfnamefont {M.}~\bibnamefont {Loidl}}, \bibinfo {author} {\bibfnamefont {X.}~\bibnamefont {Mougeot}}, \bibinfo {author} {\bibfnamefont {M.}~\bibnamefont {Paulsen}}, \bibinfo {author} {\bibfnamefont {P.}~\bibnamefont {Ranitzsch}}, \ and\ \bibinfo {author} {\bibfnamefont {M.}~\bibnamefont {Rodrigues}},\ }\bibfield  {title} {\enquote {\bibinfo {title} {High precision measurement of the $^{151}\mathrm{Sm}$ beta decay by means of a metallic magnetic calorimeter},}\ }\href {\doibase https://doi.org/10.1016/j.apradiso.2022.110237} {\bibfield  {journal} {\bibinfo  {journal} {Applied Radiation and Isotopes}\ }\textbf {\bibinfo {volume} {185}},\ \bibinfo {pages} {110237} (\bibinfo {year} {2022})}\BibitemShut {NoStop}%
\bibitem [{\citenamefont {Figueroa-Feliciano}\ \emph {et~al.}(2008)\citenamefont {Figueroa-Feliciano}, \citenamefont {Saab}, \citenamefont {Rivera-Ortiz}, \citenamefont {Bandler}, \citenamefont {Iyomoto}, \citenamefont {Kelley}, \citenamefont {Kilbourne}, \citenamefont {Porter},\ and\ \citenamefont {Smith}}]{FigueroaFeliciano2008}%
  \BibitemOpen
  \bibfield  {author} {\bibinfo {author} {\bibfnamefont {E.}~\bibnamefont {Figueroa-Feliciano}}, \bibinfo {author} {\bibfnamefont {T.}~\bibnamefont {Saab}}, \bibinfo {author} {\bibfnamefont {P.~M.}\ \bibnamefont {Rivera-Ortiz}}, \bibinfo {author} {\bibfnamefont {S.~R.}\ \bibnamefont {Bandler}}, \bibinfo {author} {\bibfnamefont {N.}~\bibnamefont {Iyomoto}}, \bibinfo {author} {\bibfnamefont {R.~L.}\ \bibnamefont {Kelley}}, \bibinfo {author} {\bibfnamefont {C.~A.}\ \bibnamefont {Kilbourne}}, \bibinfo {author} {\bibfnamefont {F.~S.}\ \bibnamefont {Porter}}, \ and\ \bibinfo {author} {\bibfnamefont {S.~J.}\ \bibnamefont {Smith}},\ }\bibfield  {title} {\enquote {\bibinfo {title} {Studies of thermal diffusion in planar absorber designs for the micro-x rocket},}\ }\href {\doibase 10.1007/s10909-007-9667-4} {\bibfield  {journal} {\bibinfo  {journal} {Journal of Low Temperature Physics}\ }\textbf {\bibinfo {volume} {151}},\ \bibinfo {pages} {424--429} (\bibinfo {year} {2008})}\BibitemShut {NoStop}%
\bibitem [{\citenamefont {Kozorezov}\ \emph {et~al.}(2013)\citenamefont {Kozorezov}, \citenamefont {Lambert}, \citenamefont {Bandler}, \citenamefont {Balvin}, \citenamefont {Busch}, \citenamefont {Nagler}, \citenamefont {Porst}, \citenamefont {Smith}, \citenamefont {Stevenson},\ and\ \citenamefont {Sadleir}}]{Kozorezov2013}%
  \BibitemOpen
  \bibfield  {author} {\bibinfo {author} {\bibfnamefont {A.~G.}\ \bibnamefont {Kozorezov}}, \bibinfo {author} {\bibfnamefont {C.~J.}\ \bibnamefont {Lambert}}, \bibinfo {author} {\bibfnamefont {S.~R.}\ \bibnamefont {Bandler}}, \bibinfo {author} {\bibfnamefont {M.~A.}\ \bibnamefont {Balvin}}, \bibinfo {author} {\bibfnamefont {S.~E.}\ \bibnamefont {Busch}}, \bibinfo {author} {\bibfnamefont {P.~N.}\ \bibnamefont {Nagler}}, \bibinfo {author} {\bibfnamefont {J.-P.}\ \bibnamefont {Porst}}, \bibinfo {author} {\bibfnamefont {S.~J.}\ \bibnamefont {Smith}}, \bibinfo {author} {\bibfnamefont {T.~R.}\ \bibnamefont {Stevenson}}, \ and\ \bibinfo {author} {\bibfnamefont {J.~E.}\ \bibnamefont {Sadleir}},\ }\bibfield  {title} {\enquote {\bibinfo {title} {Athermal energy loss from x-rays deposited in thin superconducting films on solid substrates},}\ }\href {\doibase 10.1103/PhysRevB.87.104504} {\bibfield  {journal} {\bibinfo  {journal} {Phys. Rev. B}\ }\textbf {\bibinfo {volume} {87}},\ \bibinfo {pages} {104504} (\bibinfo
  {year} {2013})}\BibitemShut {NoStop}%
\bibitem [{\citenamefont {Fleischmann}\ \emph {et~al.}(2009)\citenamefont {Fleischmann}, \citenamefont {Gastaldo}, \citenamefont {Kempf}, \citenamefont {Kirsch}, \citenamefont {Pabinger}, \citenamefont {Pies}, \citenamefont {Porst}, \citenamefont {Ranitzsch}, \citenamefont {Schäfer}, \citenamefont {Seggern}, \citenamefont {Wolf}, \citenamefont {Enss},\ and\ \citenamefont {Seidel}}]{Fleischmann2009}%
  \BibitemOpen
  \bibfield  {author} {\bibinfo {author} {\bibfnamefont {A.}~\bibnamefont {Fleischmann}}, \bibinfo {author} {\bibfnamefont {L.}~\bibnamefont {Gastaldo}}, \bibinfo {author} {\bibfnamefont {S.}~\bibnamefont {Kempf}}, \bibinfo {author} {\bibfnamefont {A.}~\bibnamefont {Kirsch}}, \bibinfo {author} {\bibfnamefont {A.}~\bibnamefont {Pabinger}}, \bibinfo {author} {\bibfnamefont {C.}~\bibnamefont {Pies}}, \bibinfo {author} {\bibfnamefont {J.}~\bibnamefont {Porst}}, \bibinfo {author} {\bibfnamefont {P.}~\bibnamefont {Ranitzsch}}, \bibinfo {author} {\bibfnamefont {S.}~\bibnamefont {Schäfer}}, \bibinfo {author} {\bibfnamefont {F.~v.}\ \bibnamefont {Seggern}}, \bibinfo {author} {\bibfnamefont {T.}~\bibnamefont {Wolf}}, \bibinfo {author} {\bibfnamefont {C.}~\bibnamefont {Enss}}, \ and\ \bibinfo {author} {\bibfnamefont {G.~M.}\ \bibnamefont {Seidel}},\ }\bibfield  {title} {\enquote {\bibinfo {title} {{Metallic magnetic calorimeters}},}\ }\href {\doibase 10.1063/1.3292407} {\bibfield  {journal} {\bibinfo  {journal} {AIP
  Conference Proceedings}\ }\textbf {\bibinfo {volume} {1185}},\ \bibinfo {pages} {571--578} (\bibinfo {year} {2009})},\ \Eprint {http://arxiv.org/abs/https://pubs.aip.org/aip/acp/article-pdf/1185/1/571/12248587/571\_1\_online.pdf} {https://pubs.aip.org/aip/acp/article-pdf/1185/1/571/12248587/571\_1\_online.pdf} \BibitemShut {NoStop}%
\bibitem [{\citenamefont {Bass}\ and\ \citenamefont {Doser}(2024)}]{Bass2024}%
  \BibitemOpen
  \bibfield  {author} {\bibinfo {author} {\bibfnamefont {S.~D.}\ \bibnamefont {Bass}}\ and\ \bibinfo {author} {\bibfnamefont {M.}~\bibnamefont {Doser}},\ }\bibfield  {title} {\enquote {\bibinfo {title} {Quantum sensing for particle physics},}\ }\href {\doibase 10.1038/s42254-024-00714-3} {\bibfield  {journal} {\bibinfo  {journal} {Nature Reviews Physics}\ }\textbf {\bibinfo {volume} {6}},\ \bibinfo {pages} {329--339} (\bibinfo {year} {2024})}\BibitemShut {NoStop}%
\bibitem [{\citenamefont {Irastorza}\ \emph {et~al.}(2013)\citenamefont {Irastorza}, \citenamefont {Armengaud}, \citenamefont {Avignone}, \citenamefont {Betz}, \citenamefont {Brax}, \citenamefont {Brun}, \citenamefont {Cantatore}, \citenamefont {Carmona}, \citenamefont {Carosi}, \citenamefont {Caspers}, \citenamefont {Caspi}, \citenamefont {Cetin}, \citenamefont {Chelouche}, \citenamefont {Christensen}, \citenamefont {Dael}, \citenamefont {Dafni}, \citenamefont {Davenport}, \citenamefont {Derbin}, \citenamefont {Desch}, \citenamefont {Diago}, \citenamefont {Dobrich}, \citenamefont {Dratchnev}, \citenamefont {Dudarev}, \citenamefont {Eleftheriadis}, \citenamefont {Fanourakis}, \citenamefont {Ferrer-Ribas}, \citenamefont {Galan}, \citenamefont {Garcia}, \citenamefont {Garza}, \citenamefont {Geralis}, \citenamefont {Gimeno}, \citenamefont {Giomataris}, \citenamefont {Gninenko}, \citenamefont {Gomez}, \citenamefont {Gonzalez-Diaz}, \citenamefont {Guendelman}, \citenamefont {Hailey}, \citenamefont {Hiramatsu},
  \citenamefont {Hoffmann}, \citenamefont {Horns}, \citenamefont {Iguaz}, \citenamefont {Isern}, \citenamefont {Imai}, \citenamefont {Jakobsen}, \citenamefont {Jaeckel}, \citenamefont {Jakovcic}, \citenamefont {Kaminski}, \citenamefont {Kawasaki}, \citenamefont {Karuza}, \citenamefont {Krcmar}, \citenamefont {Kousouris}, \citenamefont {Krieger}, \citenamefont {Lakic}, \citenamefont {Limousin}, \citenamefont {Lindner}, \citenamefont {Liolios}, \citenamefont {Luzon}, \citenamefont {Matsuki}, \citenamefont {Muratova}, \citenamefont {Nones}, \citenamefont {Ortega}, \citenamefont {Papaevangelou}, \citenamefont {Pivovaroff}, \citenamefont {Raffelt}, \citenamefont {Redondo}, \citenamefont {Ringwald}, \citenamefont {Russenschuck}, \citenamefont {Ruz}, \citenamefont {Saikawa}, \citenamefont {Savvidis}, \citenamefont {Sekiguchi}, \citenamefont {Semertzidis}, \citenamefont {Shilon}, \citenamefont {Sikivie}, \citenamefont {Silva}, \citenamefont {Kate}, \citenamefont {Tomas}, \citenamefont {Troitsky}, \citenamefont
  {Vafeiadis}, \citenamefont {van Bibber}, \citenamefont {Vedrine}, \citenamefont {Villar}, \citenamefont {Vogel}, \citenamefont {Walckiers}, \citenamefont {Weltman}, \citenamefont {Wester}, \citenamefont {Yildiz},\ and\ \citenamefont {Zioutas}}]{IAXO2013}%
  \BibitemOpen
  \bibfield  {author} {\bibinfo {author} {\bibfnamefont {I.~G.}\ \bibnamefont {Irastorza}}, \bibinfo {author} {\bibfnamefont {E.}~\bibnamefont {Armengaud}}, \bibinfo {author} {\bibfnamefont {F.~T.}\ \bibnamefont {Avignone}}, \bibinfo {author} {\bibfnamefont {M.}~\bibnamefont {Betz}}, \bibinfo {author} {\bibfnamefont {P.}~\bibnamefont {Brax}}, \bibinfo {author} {\bibfnamefont {P.}~\bibnamefont {Brun}}, \bibinfo {author} {\bibfnamefont {G.}~\bibnamefont {Cantatore}}, \bibinfo {author} {\bibfnamefont {J.~M.}\ \bibnamefont {Carmona}}, \bibinfo {author} {\bibfnamefont {G.~P.}\ \bibnamefont {Carosi}}, \bibinfo {author} {\bibfnamefont {F.}~\bibnamefont {Caspers}}, \bibinfo {author} {\bibfnamefont {S.}~\bibnamefont {Caspi}}, \bibinfo {author} {\bibfnamefont {S.~A.}\ \bibnamefont {Cetin}}, \bibinfo {author} {\bibfnamefont {D.}~\bibnamefont {Chelouche}}, \bibinfo {author} {\bibfnamefont {F.~E.}\ \bibnamefont {Christensen}}, \bibinfo {author} {\bibfnamefont {A.}~\bibnamefont {Dael}}, \bibinfo {author} {\bibfnamefont
  {T.}~\bibnamefont {Dafni}}, \bibinfo {author} {\bibfnamefont {M.}~\bibnamefont {Davenport}}, \bibinfo {author} {\bibfnamefont {A.}~\bibnamefont {Derbin}}, \bibinfo {author} {\bibfnamefont {K.}~\bibnamefont {Desch}}, \bibinfo {author} {\bibfnamefont {A.}~\bibnamefont {Diago}}, \bibinfo {author} {\bibfnamefont {B.~D.}\ \bibnamefont {Dobrich}}, \bibinfo {author} {\bibfnamefont {I.}~\bibnamefont {Dratchnev}}, \bibinfo {author} {\bibfnamefont {A.}~\bibnamefont {Dudarev}}, \bibinfo {author} {\bibfnamefont {C.}~\bibnamefont {Eleftheriadis}}, \bibinfo {author} {\bibfnamefont {G.}~\bibnamefont {Fanourakis}}, \bibinfo {author} {\bibfnamefont {E.}~\bibnamefont {Ferrer-Ribas}}, \bibinfo {author} {\bibfnamefont {J.}~\bibnamefont {Galan}}, \bibinfo {author} {\bibfnamefont {J.~A.}\ \bibnamefont {Garcia}}, \bibinfo {author} {\bibfnamefont {J.~G.}\ \bibnamefont {Garza}}, \bibinfo {author} {\bibfnamefont {T.}~\bibnamefont {Geralis}}, \bibinfo {author} {\bibfnamefont {B.}~\bibnamefont {Gimeno}}, \bibinfo {author}
  {\bibfnamefont {I.}~\bibnamefont {Giomataris}}, \bibinfo {author} {\bibfnamefont {S.}~\bibnamefont {Gninenko}}, \bibinfo {author} {\bibfnamefont {H.}~\bibnamefont {Gomez}}, \bibinfo {author} {\bibfnamefont {D.}~\bibnamefont {Gonzalez-Diaz}}, \bibinfo {author} {\bibfnamefont {E.}~\bibnamefont {Guendelman}}, \bibinfo {author} {\bibfnamefont {C.~J.}\ \bibnamefont {Hailey}}, \bibinfo {author} {\bibfnamefont {T.}~\bibnamefont {Hiramatsu}}, \bibinfo {author} {\bibfnamefont {D.~H.~H.}\ \bibnamefont {Hoffmann}}, \bibinfo {author} {\bibfnamefont {D.}~\bibnamefont {Horns}}, \bibinfo {author} {\bibfnamefont {F.~J.}\ \bibnamefont {Iguaz}}, \bibinfo {author} {\bibfnamefont {J.}~\bibnamefont {Isern}}, \bibinfo {author} {\bibfnamefont {K.}~\bibnamefont {Imai}}, \bibinfo {author} {\bibfnamefont {A.~C.}\ \bibnamefont {Jakobsen}}, \bibinfo {author} {\bibfnamefont {J.}~\bibnamefont {Jaeckel}}, \bibinfo {author} {\bibfnamefont {K.}~\bibnamefont {Jakovcic}}, \bibinfo {author} {\bibfnamefont {J.}~\bibnamefont {Kaminski}},
  \bibinfo {author} {\bibfnamefont {M.}~\bibnamefont {Kawasaki}}, \bibinfo {author} {\bibfnamefont {M.}~\bibnamefont {Karuza}}, \bibinfo {author} {\bibfnamefont {M.}~\bibnamefont {Krcmar}}, \bibinfo {author} {\bibfnamefont {K.}~\bibnamefont {Kousouris}}, \bibinfo {author} {\bibfnamefont {C.}~\bibnamefont {Krieger}}, \bibinfo {author} {\bibfnamefont {B.}~\bibnamefont {Lakic}}, \bibinfo {author} {\bibfnamefont {O.}~\bibnamefont {Limousin}}, \bibinfo {author} {\bibfnamefont {A.}~\bibnamefont {Lindner}}, \bibinfo {author} {\bibfnamefont {A.}~\bibnamefont {Liolios}}, \bibinfo {author} {\bibfnamefont {G.}~\bibnamefont {Luzon}}, \bibinfo {author} {\bibfnamefont {S.}~\bibnamefont {Matsuki}}, \bibinfo {author} {\bibfnamefont {V.~N.}\ \bibnamefont {Muratova}}, \bibinfo {author} {\bibfnamefont {C.}~\bibnamefont {Nones}}, \bibinfo {author} {\bibfnamefont {I.}~\bibnamefont {Ortega}}, \bibinfo {author} {\bibfnamefont {T.}~\bibnamefont {Papaevangelou}}, \bibinfo {author} {\bibfnamefont {M.~J.}\ \bibnamefont {Pivovaroff}},
  \bibinfo {author} {\bibfnamefont {G.}~\bibnamefont {Raffelt}}, \bibinfo {author} {\bibfnamefont {J.}~\bibnamefont {Redondo}}, \bibinfo {author} {\bibfnamefont {A.}~\bibnamefont {Ringwald}}, \bibinfo {author} {\bibfnamefont {S.}~\bibnamefont {Russenschuck}}, \bibinfo {author} {\bibfnamefont {J.}~\bibnamefont {Ruz}}, \bibinfo {author} {\bibfnamefont {K.}~\bibnamefont {Saikawa}}, \bibinfo {author} {\bibfnamefont {I.}~\bibnamefont {Savvidis}}, \bibinfo {author} {\bibfnamefont {T.}~\bibnamefont {Sekiguchi}}, \bibinfo {author} {\bibfnamefont {Y.~K.}\ \bibnamefont {Semertzidis}}, \bibinfo {author} {\bibfnamefont {I.}~\bibnamefont {Shilon}}, \bibinfo {author} {\bibfnamefont {P.}~\bibnamefont {Sikivie}}, \bibinfo {author} {\bibfnamefont {H.}~\bibnamefont {Silva}}, \bibinfo {author} {\bibfnamefont {H.~t.}\ \bibnamefont {Kate}}, \bibinfo {author} {\bibfnamefont {A.}~\bibnamefont {Tomas}}, \bibinfo {author} {\bibfnamefont {S.}~\bibnamefont {Troitsky}}, \bibinfo {author} {\bibfnamefont {T.}~\bibnamefont {Vafeiadis}},
  \bibinfo {author} {\bibfnamefont {K.}~\bibnamefont {van Bibber}}, \bibinfo {author} {\bibfnamefont {P.}~\bibnamefont {Vedrine}}, \bibinfo {author} {\bibfnamefont {J.~A.}\ \bibnamefont {Villar}}, \bibinfo {author} {\bibfnamefont {J.~K.}\ \bibnamefont {Vogel}}, \bibinfo {author} {\bibfnamefont {L.}~\bibnamefont {Walckiers}}, \bibinfo {author} {\bibfnamefont {A.}~\bibnamefont {Weltman}}, \bibinfo {author} {\bibfnamefont {W.}~\bibnamefont {Wester}}, \bibinfo {author} {\bibfnamefont {S.~C.}\ \bibnamefont {Yildiz}}, \ and\ \bibinfo {author} {\bibfnamefont {K.}~\bibnamefont {Zioutas}} (\bibinfo {collaboration} {IAXO}),\ }\href {http://cds.cern.ch/record/1567109} {\enquote {\bibinfo {title} {{The International Axion Observatory IAXO. Letter of Intent to the CERN SPS committee}},}\ }\bibinfo {type} {Tech. Rep.}\ (\bibinfo  {institution} {CERN},\ \bibinfo {address} {Geneva},\ \bibinfo {year} {2013})\BibitemShut {NoStop}%
\bibitem [{\citenamefont {Alpert}\ \emph {et~al.}(2015)\citenamefont {Alpert}, \citenamefont {Balata}, \citenamefont {Bennett}, \citenamefont {Biasotti}, \citenamefont {Boragno}, \citenamefont {Brofferio}, \citenamefont {Ceriale}, \citenamefont {Corsini}, \citenamefont {Day}, \citenamefont {De~Gerone}, \citenamefont {Dressler}, \citenamefont {Faverzani}, \citenamefont {Ferri}, \citenamefont {Fowler}, \citenamefont {Gatti}, \citenamefont {Giachero}, \citenamefont {Hays-Wehle}, \citenamefont {Heinitz}, \citenamefont {Hilton}, \citenamefont {K{\"o}ster}, \citenamefont {Lusignoli}, \citenamefont {Maino}, \citenamefont {Mates}, \citenamefont {Nisi}, \citenamefont {Nizzolo}, \citenamefont {Nucciotti}, \citenamefont {Pessina}, \citenamefont {Pizzigoni}, \citenamefont {Puiu}, \citenamefont {Ragazzi}, \citenamefont {Reintsema}, \citenamefont {Gomes}, \citenamefont {Schmidt}, \citenamefont {Schumann}, \citenamefont {Sisti}, \citenamefont {Swetz}, \citenamefont {Terranova},\ and\ \citenamefont
  {Ullom}}]{Alpert2015HOLMES}%
  \BibitemOpen
  \bibfield  {author} {\bibinfo {author} {\bibfnamefont {B.}~\bibnamefont {Alpert}}, \bibinfo {author} {\bibfnamefont {M.}~\bibnamefont {Balata}}, \bibinfo {author} {\bibfnamefont {D.}~\bibnamefont {Bennett}}, \bibinfo {author} {\bibfnamefont {M.}~\bibnamefont {Biasotti}}, \bibinfo {author} {\bibfnamefont {C.}~\bibnamefont {Boragno}}, \bibinfo {author} {\bibfnamefont {C.}~\bibnamefont {Brofferio}}, \bibinfo {author} {\bibfnamefont {V.}~\bibnamefont {Ceriale}}, \bibinfo {author} {\bibfnamefont {D.}~\bibnamefont {Corsini}}, \bibinfo {author} {\bibfnamefont {P.~K.}\ \bibnamefont {Day}}, \bibinfo {author} {\bibfnamefont {M.}~\bibnamefont {De~Gerone}}, \bibinfo {author} {\bibfnamefont {R.}~\bibnamefont {Dressler}}, \bibinfo {author} {\bibfnamefont {M.}~\bibnamefont {Faverzani}}, \bibinfo {author} {\bibfnamefont {E.}~\bibnamefont {Ferri}}, \bibinfo {author} {\bibfnamefont {J.}~\bibnamefont {Fowler}}, \bibinfo {author} {\bibfnamefont {F.}~\bibnamefont {Gatti}}, \bibinfo {author} {\bibfnamefont {A.}~\bibnamefont
  {Giachero}}, \bibinfo {author} {\bibfnamefont {J.}~\bibnamefont {Hays-Wehle}}, \bibinfo {author} {\bibfnamefont {S.}~\bibnamefont {Heinitz}}, \bibinfo {author} {\bibfnamefont {G.}~\bibnamefont {Hilton}}, \bibinfo {author} {\bibfnamefont {U.}~\bibnamefont {K{\"o}ster}}, \bibinfo {author} {\bibfnamefont {M.}~\bibnamefont {Lusignoli}}, \bibinfo {author} {\bibfnamefont {M.}~\bibnamefont {Maino}}, \bibinfo {author} {\bibfnamefont {J.}~\bibnamefont {Mates}}, \bibinfo {author} {\bibfnamefont {S.}~\bibnamefont {Nisi}}, \bibinfo {author} {\bibfnamefont {R.}~\bibnamefont {Nizzolo}}, \bibinfo {author} {\bibfnamefont {A.}~\bibnamefont {Nucciotti}}, \bibinfo {author} {\bibfnamefont {G.}~\bibnamefont {Pessina}}, \bibinfo {author} {\bibfnamefont {G.}~\bibnamefont {Pizzigoni}}, \bibinfo {author} {\bibfnamefont {A.}~\bibnamefont {Puiu}}, \bibinfo {author} {\bibfnamefont {S.}~\bibnamefont {Ragazzi}}, \bibinfo {author} {\bibfnamefont {C.}~\bibnamefont {Reintsema}}, \bibinfo {author} {\bibfnamefont {M.~R.}\ \bibnamefont
  {Gomes}}, \bibinfo {author} {\bibfnamefont {D.}~\bibnamefont {Schmidt}}, \bibinfo {author} {\bibfnamefont {D.}~\bibnamefont {Schumann}}, \bibinfo {author} {\bibfnamefont {M.}~\bibnamefont {Sisti}}, \bibinfo {author} {\bibfnamefont {D.}~\bibnamefont {Swetz}}, \bibinfo {author} {\bibfnamefont {F.}~\bibnamefont {Terranova}}, \ and\ \bibinfo {author} {\bibfnamefont {J.}~\bibnamefont {Ullom}},\ }\bibfield  {title} {\enquote {\bibinfo {title} {Holmes},}\ }\href {\doibase 10.1140/epjc/s10052-015-3329-5} {\bibfield  {journal} {\bibinfo  {journal} {The European Physical Journal C}\ }\textbf {\bibinfo {volume} {75}},\ \bibinfo {pages} {112} (\bibinfo {year} {2015})}\BibitemShut {NoStop}%
\bibitem [{\citenamefont {Brown}\ \emph {et~al.}(2008)\citenamefont {Brown}, \citenamefont {Bandler}, \citenamefont {Brekosky}, \citenamefont {Chervenak}, \citenamefont {Figueroa-Feliciano}, \citenamefont {Finkbeiner}, \citenamefont {Iyomoto}, \citenamefont {Kelley}, \citenamefont {Kilbourne}, \citenamefont {Porter}, \citenamefont {Smith}, \citenamefont {Saab},\ and\ \citenamefont {Sadleir}}]{Brown2008}%
  \BibitemOpen
  \bibfield  {author} {\bibinfo {author} {\bibfnamefont {A.-D.}\ \bibnamefont {Brown}}, \bibinfo {author} {\bibfnamefont {S.~R.}\ \bibnamefont {Bandler}}, \bibinfo {author} {\bibfnamefont {R.}~\bibnamefont {Brekosky}}, \bibinfo {author} {\bibfnamefont {J.~A.}\ \bibnamefont {Chervenak}}, \bibinfo {author} {\bibfnamefont {E.}~\bibnamefont {Figueroa-Feliciano}}, \bibinfo {author} {\bibfnamefont {F.}~\bibnamefont {Finkbeiner}}, \bibinfo {author} {\bibfnamefont {N.}~\bibnamefont {Iyomoto}}, \bibinfo {author} {\bibfnamefont {R.~L.}\ \bibnamefont {Kelley}}, \bibinfo {author} {\bibfnamefont {C.~A.}\ \bibnamefont {Kilbourne}}, \bibinfo {author} {\bibfnamefont {F.~S.}\ \bibnamefont {Porter}}, \bibinfo {author} {\bibfnamefont {S.}~\bibnamefont {Smith}}, \bibinfo {author} {\bibfnamefont {T.}~\bibnamefont {Saab}}, \ and\ \bibinfo {author} {\bibfnamefont {J.}~\bibnamefont {Sadleir}},\ }\bibfield  {title} {\enquote {\bibinfo {title} {Absorber materials for transition-edge sensor x-ray microcalorimeters},}\ }\href {\doibase
  10.1007/s10909-007-9669-2} {\bibfield  {journal} {\bibinfo  {journal} {Journal of Low Temperature Physics}\ }\textbf {\bibinfo {volume} {151}},\ \bibinfo {pages} {413--417} (\bibinfo {year} {2008})}\BibitemShut {NoStop}%
\bibitem [{\citenamefont {Irwin}\ and\ \citenamefont {Hilton}(2005)}]{Irwin2005}%
  \BibitemOpen
  \bibfield  {author} {\bibinfo {author} {\bibfnamefont {K.}~\bibnamefont {Irwin}}\ and\ \bibinfo {author} {\bibfnamefont {G.}~\bibnamefont {Hilton}},\ }\enquote {\bibinfo {title} {Transition-edge sensors},}\ in\ \href {\doibase 10.1007/10933596_3} {\emph {\bibinfo {booktitle} {Cryogenic Particle Detection}}},\ \bibinfo {editor} {edited by\ \bibinfo {editor} {\bibfnamefont {C.}~\bibnamefont {Enss}}}\ (\bibinfo  {publisher} {Springer Berlin Heidelberg},\ \bibinfo {address} {Berlin, Heidelberg},\ \bibinfo {year} {2005})\ pp.\ \bibinfo {pages} {63--150}\BibitemShut {NoStop}%
\bibitem [{\citenamefont {Bockhorn}\ \emph {et~al.}(2020)\citenamefont {Bockhorn}, \citenamefont {Paulsen}, \citenamefont {Beyer}, \citenamefont {Kossert}, \citenamefont {Loidl}, \citenamefont {N{\"a}hle}, \citenamefont {Ranitzsch},\ and\ \citenamefont {Rodrigues}}]{Bockhorn2020}%
  \BibitemOpen
  \bibfield  {author} {\bibinfo {author} {\bibfnamefont {L.}~\bibnamefont {Bockhorn}}, \bibinfo {author} {\bibfnamefont {M.}~\bibnamefont {Paulsen}}, \bibinfo {author} {\bibfnamefont {J.}~\bibnamefont {Beyer}}, \bibinfo {author} {\bibfnamefont {K.}~\bibnamefont {Kossert}}, \bibinfo {author} {\bibfnamefont {M.}~\bibnamefont {Loidl}}, \bibinfo {author} {\bibfnamefont {O.~J.}\ \bibnamefont {N{\"a}hle}}, \bibinfo {author} {\bibfnamefont {P.~C.-O.}\ \bibnamefont {Ranitzsch}}, \ and\ \bibinfo {author} {\bibfnamefont {M.}~\bibnamefont {Rodrigues}},\ }\bibfield  {title} {\enquote {\bibinfo {title} {Improved source/absorber preparation for radionuclide spectrometry based on low-temperature calorimetric detectors},}\ }\href {\doibase 10.1007/s10909-019-02274-8} {\bibfield  {journal} {\bibinfo  {journal} {Journal of Low Temperature Physics}\ }\textbf {\bibinfo {volume} {199}},\ \bibinfo {pages} {298--305} (\bibinfo {year} {2020})}\BibitemShut {NoStop}%
\bibitem [{\citenamefont {Loidl}\ \emph {et~al.}(2018)\citenamefont {Loidl}, \citenamefont {Beyer}, \citenamefont {Bockhorn}, \citenamefont {Enss}, \citenamefont {Gy{\"o}ri}, \citenamefont {Kempf}, \citenamefont {Kossert}, \citenamefont {Mariam}, \citenamefont {N{\"a}hle}, \citenamefont {Paulsen}, \citenamefont {Rodrigues},\ and\ \citenamefont {Schmidt}}]{Loidl2018}%
  \BibitemOpen
  \bibfield  {author} {\bibinfo {author} {\bibfnamefont {M.}~\bibnamefont {Loidl}}, \bibinfo {author} {\bibfnamefont {J.}~\bibnamefont {Beyer}}, \bibinfo {author} {\bibfnamefont {L.}~\bibnamefont {Bockhorn}}, \bibinfo {author} {\bibfnamefont {C.}~\bibnamefont {Enss}}, \bibinfo {author} {\bibfnamefont {D.}~\bibnamefont {Gy{\"o}ri}}, \bibinfo {author} {\bibfnamefont {S.}~\bibnamefont {Kempf}}, \bibinfo {author} {\bibfnamefont {K.}~\bibnamefont {Kossert}}, \bibinfo {author} {\bibfnamefont {R.}~\bibnamefont {Mariam}}, \bibinfo {author} {\bibfnamefont {O.}~\bibnamefont {N{\"a}hle}}, \bibinfo {author} {\bibfnamefont {M.}~\bibnamefont {Paulsen}}, \bibinfo {author} {\bibfnamefont {M.}~\bibnamefont {Rodrigues}}, \ and\ \bibinfo {author} {\bibfnamefont {M.}~\bibnamefont {Schmidt}},\ }\bibfield  {title} {\enquote {\bibinfo {title} {Metrobeta: Beta spectrometry with metallic magnetic calorimeters in the framework of the european program of ionizing radiation metrology},}\ }\href {\doibase 10.1007/s10909-018-1933-0}
  {\bibfield  {journal} {\bibinfo  {journal} {Journal of Low Temperature Physics}\ }\textbf {\bibinfo {volume} {193}},\ \bibinfo {pages} {1251--1256} (\bibinfo {year} {2018})}\BibitemShut {NoStop}%
\bibitem [{\citenamefont {Schneider}\ \emph {et~al.}(2016)\citenamefont {Schneider}, \citenamefont {Chrysalidis}, \citenamefont {Dorrer}, \citenamefont {Düllmann}, \citenamefont {Eberhardt}, \citenamefont {Haas}, \citenamefont {Kieck}, \citenamefont {Mokry}, \citenamefont {Naubereit}, \citenamefont {Schmidt},\ and\ \citenamefont {Wendt}}]{Schneider2016}%
  \BibitemOpen
  \bibfield  {author} {\bibinfo {author} {\bibfnamefont {F.}~\bibnamefont {Schneider}}, \bibinfo {author} {\bibfnamefont {K.}~\bibnamefont {Chrysalidis}}, \bibinfo {author} {\bibfnamefont {H.}~\bibnamefont {Dorrer}}, \bibinfo {author} {\bibfnamefont {C.}~\bibnamefont {Düllmann}}, \bibinfo {author} {\bibfnamefont {K.}~\bibnamefont {Eberhardt}}, \bibinfo {author} {\bibfnamefont {R.}~\bibnamefont {Haas}}, \bibinfo {author} {\bibfnamefont {T.}~\bibnamefont {Kieck}}, \bibinfo {author} {\bibfnamefont {C.}~\bibnamefont {Mokry}}, \bibinfo {author} {\bibfnamefont {P.}~\bibnamefont {Naubereit}}, \bibinfo {author} {\bibfnamefont {S.}~\bibnamefont {Schmidt}}, \ and\ \bibinfo {author} {\bibfnamefont {K.}~\bibnamefont {Wendt}},\ }\bibfield  {title} {\enquote {\bibinfo {title} {Resonance ionization of holmium for ion implantation in microcalorimeters},}\ }\href {\doibase https://doi.org/10.1016/j.nimb.2015.12.012} {\bibfield  {journal} {\bibinfo  {journal} {Nuclear Instruments and Methods in Physics Research Section B:
  Beam Interactions with Materials and Atoms}\ }\textbf {\bibinfo {volume} {376}},\ \bibinfo {pages} {388--392} (\bibinfo {year} {2016})},\ \bibinfo {note} {proceedings of the XVIIth International Conference on Electromagnetic Isotope Separators and Related Topics (EMIS2015), Grand Rapids, MI, U.S.A., 11-15 May 2015}\BibitemShut {NoStop}%
\bibitem [{\citenamefont {Niemeyer}\ \emph {et~al.}(tion)\citenamefont {Niemeyer}, \citenamefont {Mowitz}, \citenamefont {Berndt}, \citenamefont {Beyer}, \citenamefont {Dorrer}, \citenamefont {Düllmann}, \citenamefont {Göggelmann}, \citenamefont {Hasse}, \citenamefont {Kempf}, \citenamefont {Kieck}, \citenamefont {Kneip}, \citenamefont {Kossert}, \citenamefont {Mokry}, \citenamefont {Müller}, \citenamefont {Nähle}, \citenamefont {Renisch}, \citenamefont {Runke}, \citenamefont {Studer}, \citenamefont {Takács},\ and\ \citenamefont {Wendt}}]{Niemeyer2024}%
  \BibitemOpen
  \bibfield  {author} {\bibinfo {author} {\bibfnamefont {T.}~\bibnamefont {Niemeyer}}, \bibinfo {author} {\bibfnamefont {D.}~\bibnamefont {Mowitz}}, \bibinfo {author} {\bibfnamefont {S.}~\bibnamefont {Berndt}}, \bibinfo {author} {\bibfnamefont {J.}~\bibnamefont {Beyer}}, \bibinfo {author} {\bibfnamefont {H.}~\bibnamefont {Dorrer}}, \bibinfo {author} {\bibfnamefont {C.~E.}\ \bibnamefont {Düllmann}}, \bibinfo {author} {\bibfnamefont {A.}~\bibnamefont {Göggelmann}}, \bibinfo {author} {\bibfnamefont {R.}~\bibnamefont {Hasse}}, \bibinfo {author} {\bibfnamefont {S.}~\bibnamefont {Kempf}}, \bibinfo {author} {\bibfnamefont {T.}~\bibnamefont {Kieck}}, \bibinfo {author} {\bibfnamefont {N.}~\bibnamefont {Kneip}}, \bibinfo {author} {\bibfnamefont {K.}~\bibnamefont {Kossert}}, \bibinfo {author} {\bibfnamefont {C.}~\bibnamefont {Mokry}}, \bibinfo {author} {\bibfnamefont {M.}~\bibnamefont {Müller}}, \bibinfo {author} {\bibfnamefont {O.~J.}\ \bibnamefont {Nähle}}, \bibinfo {author} {\bibfnamefont {D.}~\bibnamefont
  {Renisch}}, \bibinfo {author} {\bibfnamefont {J.}~\bibnamefont {Runke}}, \bibinfo {author} {\bibfnamefont {D.}~\bibnamefont {Studer}}, \bibinfo {author} {\bibfnamefont {M.~P.}\ \bibnamefont {Takács}}, \ and\ \bibinfo {author} {\bibfnamefont {K.}~\bibnamefont {Wendt}},\ }\bibfield  {title} {\enquote {\bibinfo {title} {Ion implantation of the electron-capture nuclide $^{55}$fe for measurements by means of metallic microcalorimeters},}\ }\href@noop {} {\  (\bibinfo {year} {in preparation})}\BibitemShut {NoStop}%
\bibitem [{\citenamefont {Friedrich}, \citenamefont {Boyd},\ and\ \citenamefont {Cantor}(2016)}]{Friedrich2016}%
  \BibitemOpen
  \bibfield  {author} {\bibinfo {author} {\bibfnamefont {S.}~\bibnamefont {Friedrich}}, \bibinfo {author} {\bibfnamefont {S.}~\bibnamefont {Boyd}}, \ and\ \bibinfo {author} {\bibfnamefont {R.}~\bibnamefont {Cantor}},\ }\bibfield  {title} {\enquote {\bibinfo {title} {Fabrication of gamma detectors based on magnetic ag:er microcalorimeters},}\ }\href {\doibase 10.2172/1256452} {\  (\bibinfo {year} {2016}),\ 10.2172/1256452}\BibitemShut {NoStop}%
\bibitem [{\citenamefont {Yagi}\ \emph {et~al.}(2023)\citenamefont {Yagi}, \citenamefont {Hayashi}, \citenamefont {Tanaka}, \citenamefont {Miyagawa}, \citenamefont {Ota}, \citenamefont {Yamasaki}, \citenamefont {Mitsuda}, \citenamefont {Yoshida}, \citenamefont {Saito},\ and\ \citenamefont {Homma}}]{Yagi2023}%
  \BibitemOpen
  \bibfield  {author} {\bibinfo {author} {\bibfnamefont {Y.}~\bibnamefont {Yagi}}, \bibinfo {author} {\bibfnamefont {T.}~\bibnamefont {Hayashi}}, \bibinfo {author} {\bibfnamefont {K.}~\bibnamefont {Tanaka}}, \bibinfo {author} {\bibfnamefont {R.}~\bibnamefont {Miyagawa}}, \bibinfo {author} {\bibfnamefont {R.}~\bibnamefont {Ota}}, \bibinfo {author} {\bibfnamefont {N.~Y.}\ \bibnamefont {Yamasaki}}, \bibinfo {author} {\bibfnamefont {K.}~\bibnamefont {Mitsuda}}, \bibinfo {author} {\bibfnamefont {N.}~\bibnamefont {Yoshida}}, \bibinfo {author} {\bibfnamefont {M.}~\bibnamefont {Saito}}, \ and\ \bibinfo {author} {\bibfnamefont {T.}~\bibnamefont {Homma}},\ }\bibfield  {title} {\enquote {\bibinfo {title} {Fabrication of a 64-pixel tes microcalorimeter array with iron absorbers uniquely designed for 14.4-kev solar axion search},}\ }\href {\doibase 10.1109/TASC.2023.3254488} {\bibfield  {journal} {\bibinfo  {journal} {IEEE Transactions on Applied Superconductivity}\ }\textbf {\bibinfo {volume} {33}},\ \bibinfo {pages}
  {1--5} (\bibinfo {year} {2023})}\BibitemShut {NoStop}%
\end{thebibliography}%


%

\end{document}